\documentclass{article}
\usepackage{arxiv}

\usepackage[utf8]{inputenc} 
\usepackage[T1]{fontenc}    
\usepackage{hyperref}       
\usepackage{url}            
\usepackage{booktabs}       
\usepackage{amsfonts}       
\usepackage{nicefrac}       
\usepackage{microtype}      
\usepackage{lipsum}
\usepackage{authblk}
\usepackage{graphicx}
\usepackage{subcaption}
\usepackage{comment}
\usepackage{appendix}
\usepackage{longtable}
\graphicspath{ {./images/} }

\title{AI Specialization for Pathways of \\
Economic Diversification}

\makeatletter
\renewcommand\AB@affilsepx{, \protect\Affilfont}
\makeatother
\author[1]{Saurabh Mishra}
\author[2]{Robert Koopman}
\author[3]{Giuditta De-Prato} 
\author[4]{Anand Rao} 
\author[5]{Israel Osorio-Rodarte}
\author[6]{Julie Kim} 
\author[7]{Nikola Spatafora} 
\author[8]{Keith Strier} 
\author[9]{Andrea Zaccaria}

\affil[1]{Institute for Human-Centered Artificial Intelligence (HAI), Stanford University} 
\affil[2]{World Trade Organization (WTO)}
\affil[3]{Joint Research Center (JRC), European Commission}
\affil[4]{PricewaterhouseCoopers (PwC)}
\affil[5]{World Bank Group}
\affil[6]{Netbase Quid}
\affil[7]{International Monetary Fund (IMF)}
\affil[8]{NVIDIA Corp.}
\affil[9]{Institute for Complex Systems, CNR}

\begin{document}
\maketitle

\begin{abstract}
The growth in AI is rapidly transforming the structure of economic production. However, very little is known about how within-AI specialization may relate to broad-based economic diversification. This paper provides a data-driven framework to integrate the interconnection between AI-based specialization with goods and services export specialization to help design future comparative advantage based on the inherent capabilities of nations. Using detailed data on private investment in AI and export specialization for more than 80 countries, we propose a systematic framework to help identify the connection from AI to goods and service sector specialization. The results are instructive for nations that aim to harness AI specialization to help guide sources of future competitive advantage.  The operational framework could help inform the public and private sector to uncover connections with nearby areas of specialization.
\end{abstract}


\newpage
\section{Introduction}

Artificial intelligence (AI) carries the promise of making industry more efficient and our lives easier. AI has been called the “new electricity” \cite{ng2018ai} reflecting an economic framing of AI as a “general purpose technology” (GPT) \cite{brynjolfsson2021productivity}. AI may be poised to join the steam engine, the electric motor, and the silicon wafer as a “technological prime mover”, \cite{bresnahan1995general} but it may also might be "disruptive" and cause structural changes (which then often generates resistance by some vested interests). With this promise, however, also comes the fear of job replacement, manufacturing stagnation, hollowing out of the middle class, increased income inequality \cite{goldin2020technology, korinek2021artificial, huang2018artificial}. With the growing call for national AI strategies, governments around the world have a mandate to leverage AI for economic competitiveness and to help shape their industrial strategy. Rather than looking at AI as a potential threat, AI can be viewed as one factor in economic transformation. However, very little is known about sources of AI specialization, and how specialization in AI could be leveraged for strategic diversification opportunities based on existing endowments of a country. How might developing countries leverage AI to keep traditional industries competitive and discover new opportunities?

Take the example of a country like Bangladesh, an economy that is characterized by low-diversification and a specialization in the garments industry. How could AI help Bangladesh achieve more prosperity for its citizens? This question can be viewed from different lenses. For example, which AI use-cases are most closely related to the garment industry and may be deployed for competitive advantage? Identifying sub-sectors in AI  could help efficiency of Bangladesh's garment production chain and create easier pathway to discover new specializations most aligned with garment and related-AI capabilities.\footnote{Low-income or lower-middle income countries like Bangladesh may not have the capability to produce certain type of AI domestically. However, importing strategic AI sectors may help  traditional industries become more competitive which in turn could help discover pathways for developing capabilities in other AI, goods, or services.} Second, how might any existing capabilities in Bangladesh's AI ecosystem create opportunities for broad-based economic diversification i.e. discovery of specialization in nearby goods and services? Third, how could existing specializations in goods and services lead to developing new AI capabilities for Bangladesh's economy? Answers to these fundamental questions can offer pathways for developing countries to formulate data-driven industrial and growth strategies that promote diversification and long-term prosperity. 

Some of the more popular AI models are large neural networks that run on state-of-the-art machines. For example, AlphaGo Zero and GPT-3 require millions of dollars in computing power for training AI research and AI research is being shaped by a few actors, who are mostly affiliated with either large technology firms or elite universities \cite{ahmed2020democratization}. These growing fears of missing out (FOMO) has led many to believe that AI may leave developing countries behind. Progress in AI is being driven by the availability of a large volume of training data, algorithmic innovations, and computing capabilities. For example, the time to train object detection task like ImageNet to over 90\% accuracy has been  reduced from over 10 hours to a few seconds, and the cost has declined from over \$2,000 to substantially less than \$10 within the span of the last three years. Leveraging capabilities in AI will be pivotal for competitiveness of traditional industries and discovery of new industries, especially for emerging markets (EM’s). 

This paper measures cross-country specialization in AI, to inform how current AI specializations could guide pathways for future diversification. The paper addresses three fundamental questions: (1) what is the optimal way to measure AI specialization inclusive of economic specialization patterns, not just in manufacturing and goods but also service specialization? (2) where does AI specialization appear to fit in to global production networks --- are they central or peripheral, i.e., what are the backward and forward linkages and plausibility of knowledge spillovers between AI services and other sectors of the economy? and (3) might AI specialization provide pathways for new applications to integrate inherent capabilities of an economy’s production pattern? These questions are important for advanced economies too but particularly for developing countries' with younger industrial strategies. Developing countries are especially vulnerable to poorly constructed national AI plans. These often default towards “vanity” projects at the expense of more sophisticated investments that capitalize on AI specialization and promote diversification without accounting for core areas of economic specialization.  It is also common to find discussions in military and defense circles on technology offsets i.e., combining various technologies to build a new source of specialization to drive growth and competitiveness.	
		
Pathways for economic diversification have emerged as a central theme in economic growth and development literature. A nascent but growing body of literature provides insights into structural transformation, exports and development. Economic diversification is known to follow a non-monotonic path over the stages of economic development \cite{imbs2003stages, henn2013export, henn2020export} and there is a positive relationship between structural shifts and per capita income \cite{pineres2000commodity, herzer2006export, al2000export, de2002natural, cadot2011export}. Diversification of exports (in terms of products and destinations) helps to stabilize export earnings in the longer run, with benefits analogous to the portfolio effect in finance \cite{ghosh1994export, bleaney2001impact}. Recent work has accounted for combining goods and services specialization to offer a more comprehensive view of national diversification patterns \cite{stojkoski2016impact,zaccaria2018integrating, mishra2020economic}. However, in the economic and policy literature, AI has been missing from economic diversification discussions. 	

AI is a data-driven field but, paradoxically, detailed data about within-AI specialization at a cross-country level is limited \cite{index2019annual}. Beyond data limitations, there are measurement challenges, especially those regarding consistent ways to quantify AI’s technical progress, investment, or research and development output across countries. For example, national statistics agencies methodologies do not measure public investments in AI R\&D or private investment in AI in a similar manner to classical sectors \cite{mishra2020measurement}. Unlike data for economic statistics such as goods exports or services exports, that follows international classification standards, for example, SITC, HS, ISIC, etc., data on AI is not collected in standard format from statistical agencies. Data on corporate activity collected through digital channels is potentially a formidable opportunity for empirical researchers and has the potential to significantly improve our ability to track changes in firms’ technological configurations \cite{horton2015labor}. Since several disparate inputs are required to put AI into production — human capital, software, data, computational power, and new management practices — none of these are easy to observe and statistical agencies do not regularly collect data on them. 

As firms are still adjusting inputs to AI and experimenting with the technology, investments data tends to be a noisy indicator. For instance, rapid changes to AI technologies can change the value of investments in AI-related skills \cite{rock2019engineering}. Illustratively, what was a hot skill may become a commodity-skill two years later. It is important to measure AI inputs, i.e., skills, software, data, and management practices as well as AI outputs, for example, user consumption patterns related to AI products and services. Both are difficult to measure, however, as they are often service-based, and their intangibility remains a measurement challenge \cite{mishra2020measurement}. All of these challenges are amplified when assessing developing countries. To address these challenges at the firm-level, we use macro-data on private sector AI investments from credible and reliable sources, such as Crunchbase, CapIQ, Netbase Quid, to serve as proxies to measure detailed sub-sectors of private investment in AI at the national level. 

To aid potential pathways for economic diversification in emerging markets (EM’s), the field of economic complexity \cite{hidalgo2021economic} has emerged as an important metric to measure nations’ inherent capabilities embodied in the structure of economic production. Acquiring the knowledge to host these capabilities — physical and human capital, institutions, organizational abilities— are reflected in the ability to produce and export sophisticated products and services \cite{hausmann2007you, hidalgo2009building, mishra2020economic,zaccaria2018integrating,cristelli2013measuring}; as a consequence, the co-occurrence of products in countries and firms can be used to reconstruct the relatedness of industrial, technological, and scientific sectors \cite{teece1994understanding,hidalgo2018principle,zaccaria2014taxonomy,hidalgo2007product,pugliese2019unfolding} .
In similar vein, the trends of investment in specific AI sub-sectors offers a proxy of capabilities to produce specific AI specializations that are present in the startup community and are attracting those investments. Accounting for AI specialization along with broader economic specialization remains pivotal for EM’s that are unable to generate sufficiently high-growth and could get stuck in a “middle-income” trap or a “resource-curse” with limited opportunities for diversification \cite{battaile2015transforming, felipe2012tracking, flaaen2013avoid}. EM’s can leverage specific AI specializations for transforming their production pattern to help make pre-existing industries more competitive or discover new sources of comparative advantage. In this paper we aim at applying the economic complexity methodology to connect specific AI sectors to goods and service exports and to other AI sectors. In particular, we compute statistically validated normalized co-occurrences of trade and AI sectors in countries, taking into account time evolution in an explicit way and thus providing an arrow of development from AI sectors to trade sectors. This will allow us to build a recommendation framework to detect which AI investment better supports a specific target industry or service.

The rest of the paper is organized as follows: Section 2 outlines the data and methods. Section 3 presents the results from the global network of AI to goods and services specialization. Section 4 discusses an operational framework to help guide country specific diversification strategies. Section 5 concludes with applications of our framework and future work.

\section{Data and Methods}
\label{sec:data}

\subsection{Data}
This section presents the diverse sources of data and our proposed novel methodology to link AI specializations with sources of comparative advantage in goods and services. Traditionally, economic complexity analysis has been conducted with goods exports statistics \cite{hausmann2007you, hidalgo2007product, cristelli2013measuring}, while integration with service database is relatively recent \cite{stojkoski2016impact,zaccaria2018integrating, mishra2020economic}. The trade data is from U.N. COMTRADE (\url{https://comtrade.un.org}) for physical goods and from the Balance of Payments database collected by the IMF (\url{https://data.imf.org}). This two databases both refer to export volumes, so they can be integrated giving rise to what we call the Universal database \cite{zaccaria2018integrating}. On the other hand, AI specialization is largely missing in complex network studies. To study comparative advantage in specific AI sectors, we use data from diverse sources including private investment in AI; in particular, we use data on AI investment from Crunchbase, CAPIQ, Netbase Quid that originally provides data for 49 AI sub-sectors that we manually curate to obtain 39 AI sub-sectors \footnote{Quid embeds organizational data from CapIQ and Crunchbase. More details on the AI investment data can be found in \nameref{sec:appendix}. The AI private investment is comprehensive and accurate source that embeds investments from global organizations, including early-stage startups and funding events data providing a strong signal of AI related investment activities. Quid then uses boolean query to search for focus areas, keywords within the companies database, and categorizes and visualizes these companies based on their business descriptions using Quid’s proprietary NLP algorithm. More details on the proprietary algorithm can be found at \href{https://netbasequid.com/applications/quid-technologies}{Quid technologies}. The Quid methodology is also available from the \href{https://hai.stanford.edu}{Stanford Institute for Human-Centered Artificial Intelligence} AI Index Report \href{https://aiindex.stanford.edu/wp-content/uploads/2021/03/2021-AI-Index-Report-_Chapter-3.pdf}{Chapter 3 (page 31-32)}.}. We index 3.6M public and private company profiles from multiple data sources that are indexed to search across company descriptions, while filtering and including metadata ranging from investment information to firmographic information such as year founded, head quarters (HQ) location, and more. Company information is updated on a weekly basis. Trends are based on reading the text description of companies to identify key words, phrases, people, companies, and institutions; then comparing different words from each document (news article, company descriptions…etc) to develop links between these words based on similar language. This process is repeated at immense scale which produces a network that shows how similar all documents are. AI investment is concentrated in select countries, for example, graphs in the \nameref{sec:appendix} show that between 2019-2020, North America accounted for for 58\% of global AI investment, followed by East Asia \& Pacific (28\%), Europe and Central Asia (10\%), Middle-East and North Africa (2\%), South Asia (1\%), Latin America (0.5\%), and Sub-Saharan Africa (0.2\%). China alone accounts for approximately 20\% of global AI investment in 2019-2020. Details about the data, sources, and definitions about AI,  goods and services are available in the \nameref{sec:appendix}. The universal AI dataset, represented as UAI incorporates the AI matrix denoted as (A), goods matrix denoted as (G), and the services matrix denoted as (S). This is summarized in Table \ref{tab:table1}.

\begin{table}
[ht]
 \caption{Nomenclature for AI (A), Goods (G), Services (S)  and Universal (UAI) datasets}
  \centering
  \begin{tabular}{lcc}
    \toprule
    \cmidrule(r){1-2}
    Sector     &      Nomenclature   \\
    \midrule
    Goods (or Products) & G       \\
    Services     & S       \\
    Artificial Intelligence     & AI      \\
     Universal Artificial Intelligence     & UAI      \\
    \bottomrule
  \end{tabular}
  \label{tab:table1}
\end{table}

\subsection{Methods}
\subsubsection{Detecting specialization dynamics}

In order to detect whether we can say that a country is specialized in a specific trade (G or S) or AI sector, we first compute the Revealed Comparative Advantage (RCA), that informs whether a country’s share of an item in it's export (or investment) basket is larger (or smaller) than the items’ share in the world export (or investment) basket \cite{balassa1965trade}, where an item can be an AI, an industrial, or a service sub-sector.\footnote{Note that a country could have a RCA higher than 1 in an AI sub-sector if it invested even a small amount in the one AI sub-sector, and nothing at all in any of the other AI sub-sectors. The advantage is that RCA will provide a clear and properly normalized view of the relative specialization. This is the well-recognized method to obtain a natural threshold of relative specialization, not just in exports but also in scientific and patenting activities \cite{hidalgo2007product, cristelli2013measuring, hidalgo2021economic, zaccaria2018integrating, pugliese2019unfolding}.} The RCAs are then converted to binary values (i.e., 0/1) based on whether they are above a threshold or below it, where the generally accepted threshold is 1. The binary matrix is used to compute the Assist Matrix, which is then statistically validated. The RCA matrices are computed independently for AI (A), Goods (G), and Service (S) matrices to create a dataset for universal AI (UAI) that is used to build the global progression network from AI to broader trade networks. 

In formula, RCA of country $c$ in item $x$ is computed as:

\begin{equation}
\mathrm{RCA}_{c,x}=\frac {E _{c,x}/\sum_x E _{c,x}}{\sum_c E _{c,x}/\sum_{c,x} E _{c,x}}
\end{equation}

where $x$ is either $\{a,g,s\}$, that is AI, goods (or products), or services sectors, and $E_{c,x}$ is the value of products $g$ or services $s$, in US dollars, exported by country $c$; if instead the RCA is computed for the AI investment data, $E_{c,a}$ is the investment of country $c$ in the AI sector $a$. We refer to items (or activities) as individual categories (or sub-sectors) within AI investment, goods exports, or services exports.

Then, the ${c,x}$ element of the binary matrix $\textbf{M}$ is defined to be $1$ if $\mathrm{RCA}_{c,x}>1$ and zero otherwise. Note that all these quantities are time-dependent and that we have dropped the temporal variable $t$ only to obtain a lighter notation. However, the key element of our analysis is actually to compare $\textbf{M}(t)$ with $\textbf{M}(t+\Delta t)$, in order to study how an AI investment of a country influences its products or services trade specialization.

In addition, we examine the evolution of AI (A), Goods (G), and Services (S) from a slightly different viewpoint, analyzing the specialization dynamics of countries, that can consist in focusing on traditional items, or engaging in completely new economic activities. To this end, specialization patterns are classified into four categories: “Classic”, “Absent”, “Disappearing”, and “Emerging”. A “classic” specialization is defined as an item in which a country had, on average, a Revealed  Comparative Advantage (RCA) in both the 2010–14 and 2017–19 sub-periods \cite{anand2015make}. In other words, the average relative specialization of an item (whether in AI, goods, or services) remained above 1 in the earlier and later sample period. “Absent” specializations are those in which the country never had and average RCA>1. “Disappearing” specialization are those in which a country had an average RCA>1 at the initial period, but not at the final period. Conversely, “Emerging” specialization are those in which a country only developed RCA>1 at the end of the sample period.  Table \ref{tab:table2} summarizes these definitions.

\begin{table}
[ht!]
 \caption{Labelling Specialization Patterns across AI (A), Goods (G), and Services (S)}
  \centering
  \begin{tabular}{lcc}
    \toprule
    \cmidrule(r){1-2}
    Label     &      Binary RCA in 2010-14   &      Binary RCA in 2017-19   \\
    \midrule
   Classic & 1  & 1    \\
    Absent     & 0 & 0     \\
    Disappearing     & 1 & 0      \\
    Emerging     & 0 & 1       \\
    \bottomrule
  \end{tabular}
  \label{tab:table2}
\end{table}

\subsubsection{Progression Network}

This section introduces our methodology --- the Progression Network --- to estimate the weight of the connections from AI sectors to trade sectors or other AI sectors. This technique has been introduced by \cite{pugliese2019unfolding}, adopted for the universal database (goods and services) in \cite{zaccaria2018integrating} and is inspired by the Assist Matrix introduced in \cite{zaccaria2014taxonomy} and the recommendation system discussed in \cite{zhou2007bipartite}. 

\begin{figure}
    \centering
    \includegraphics[height=0.35\paperheight]{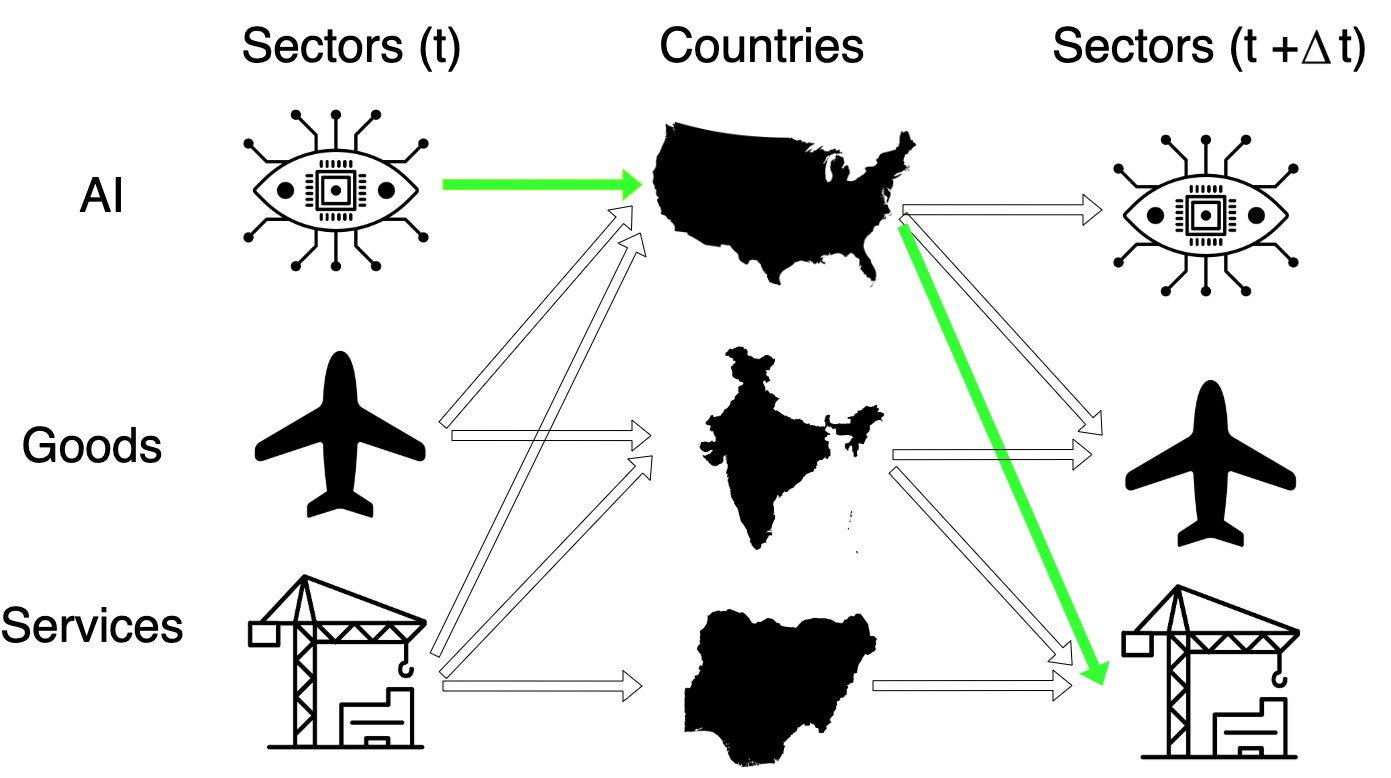}
    \caption{A schematic representation of the connections among sectors via the respective countries specialization. This idea is at the basis of the algorithm to build the progression network, as explained in the main text. The bipartite networks connecting  G, S, and AI with countries are projected to new bipartite or monopartite networks, connecting for instance AI with G and S.}
    \label{fig:abstraction1}
\end{figure}

Consider the schematic representation of Figure~\ref{fig:abstraction1}. Each arrow represents a nonzero element of the export or investment matrix $\mathbf{M}$ discussed previously: for instance, the fluorescent arrows represent USA having a Revealed Comparative Advantage (RCA) higher than one in computer vision (AI) at time $t$ and in the export of construction (Services) at time $t$ plus a given time interval $\Delta$, thus introducing a delay (in the following we will let this time delay vary). In this hypothetical example, when a country at time $t$ specializes in computer vision, and after three years it will specialize in construction --- if this pattern is systematically present in many countries, then there appears to be a \textit{connection}, a directed link, going from computer vision to construction which implies a time-delayed correlation between these two industrial sectors (from computer vision in AI to construction in services). The network, nodes, and statistically validated links between those sectors, form the idea of the the Progression Network (PN). More specifically, we will estimate the conditional probability that a bit of information, modeled to be a random walker in the tripartite UAI-Countries-UAI network, starting from a specific AI sector will arrive to a specific AI (A), goods (G), or services (S) sector. This probability is given by the Assist Matrix $\mathbf{B}(t,t+\Delta t)$, whose elements are

\begin{equation}
\label{assist}
B_{x,x'}(y,y+\Delta t)=\frac{1}{u_{x}(y)}\sum_{c}
\frac{M_{c,x}(y)M_{c,x'}(y+\Delta t)}{d_c(y+\Delta)}
\end{equation}

where $x$ is the starting or source AI sector, while $x'$ is the final or target trade or AI sector (that is, $x={a}$ and $x'={a',g,s}$). Note that the two normalizations, the first using the ubiquity of the source sector $u_x(t)=\sum_c M_{c,x}(t)$ and the second using the diversification of the countries in the middle layer $d_c(t+\Delta t)=\sum_x M_{c,x}(t+\Delta t)$, take into account the number of possible paths this bit of information may take. The economic interpretation is that if a country is very diversified, like the USA, the information coming from this country is less, and the same is valid for goods, services, or AI that are more ubiquitous.

The matrix $\mathbf{B}$ can be seen as the adjancency matrix of the network defined by the collection of nodes (the UAI sectors) and the directed links, or arrows (since $\mathbf{B}$ is not symmetrical), among them. This network is almost completely connected, in the sense that practically all the possible links are present. However, we need to discriminate the real inter-dependencies from the random co-occurrences coming simply from the fact that a product is very common, or a country is highly diversified. To do so we \textit{statistically validate} each link by comparing its weight with an ensemble of matrices generated using a null model, the Biparite Configuration Model \cite{saracco2015randomizing,saracco2017inferring}. The idea, borrowing concepts and methods from statistical physics, is to fix some constraints (in this case, we fix the average ubiquity and diversification) and to randomize everything else \cite{cimini2019statistical}, generating in such a way an ensemble of null matrices $\mathbf{\tilde{B}}$ whose elements will produce a probability density distribution for each link ${x,x'}$. Each element of the empirical matrix $\mathbf{B}$ will be compared with the respective distribution generated using the null model; if the weight is larger than a given percentile then the link from $x$ to $x'$ will be statistically validated.

This procedure is repeated for all the years $(t,t+\Delta t)$ in the database. In order to obtain a comprehensive view of the global connections among the sectors, that is a unique, time independent matrix $\mathbf{B}$, we selected only those links that have been statistically validated with pvalue 0.05; as weights, we considered the average weights of the corresponding validated links.

\begin{figure}
[ht]
    \centering
    \includegraphics[height=0.35\paperheight]{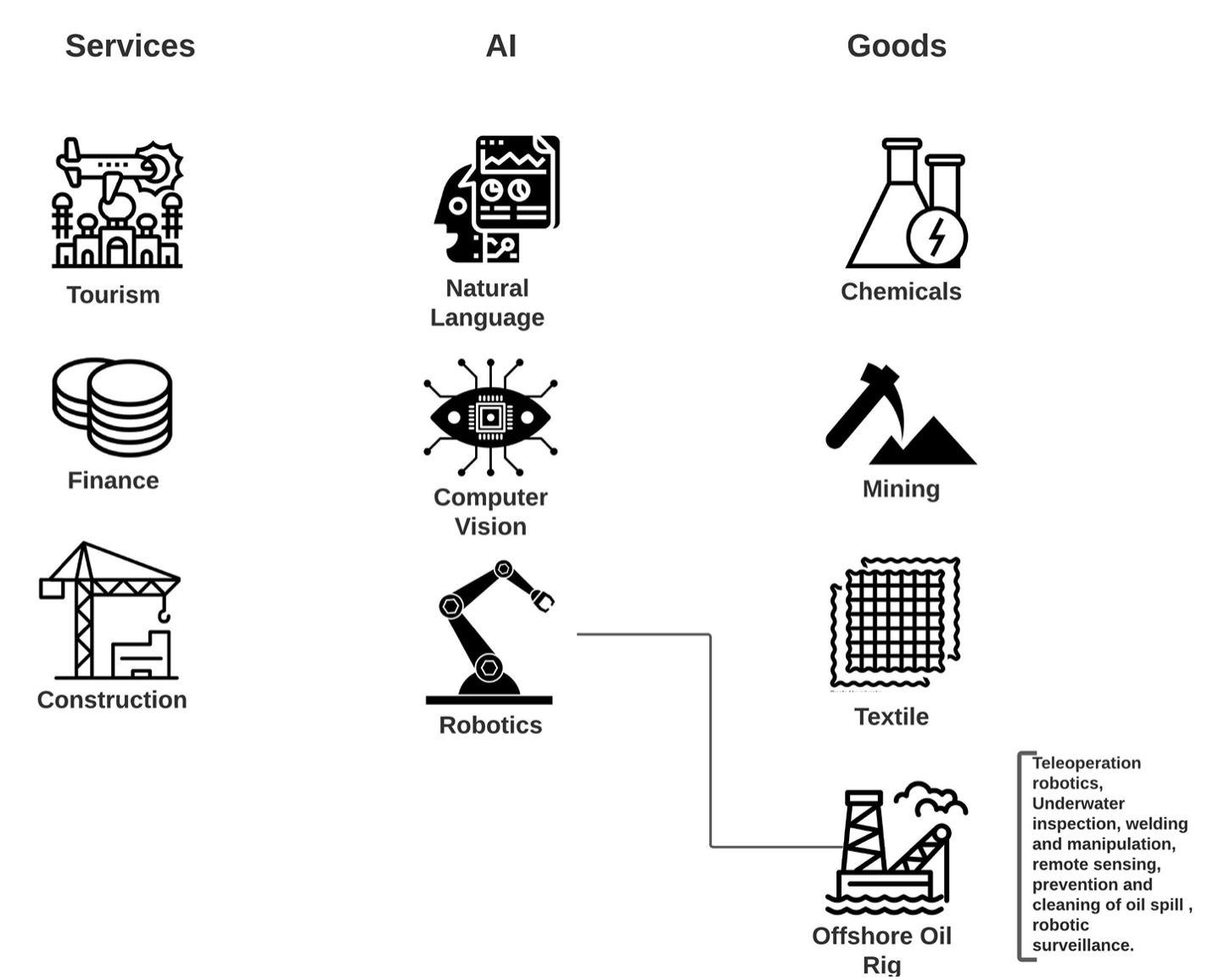}\newline
    \caption{Sample Abstraction for Linkages in a Country’s Specialization Pattern.}
    \label{fig:abstraction2}
\end{figure}

To make the connection between varied sectors within AI, goods, or services more practical, Figure~\ref{fig:abstraction2} presents an operational abstraction. Take for example an economy like Bulgaria or Jordan. The data shows that there is growing investment and specialization in these countries in Robotic Process Automation (AI) and offshore oil rig (Goods). Given the specialization in these co-occurring sectors, there may be opportunities to connect them where robotic capabilities can be applied to offshore oil drilling for making the oil industry more productive and competitive, with applications such as teleoperation robotics, underwater inspection, or robotic surveillance. Another line of reasoning could be a reverse linkage, where there is specialization in offshore oil rig but not in Robotic Automation. Given the learning from the global network that informs the strong linkage between these two sectors, either the country could import Robotic Automation capabilities or through connection with other AI and broader economic specializations help create a pathway to develop specialization internally in Robotic Automation. 

\subsubsection{Density computation}
Once the connections $\mathbf{B}$ from AI sectors to trade are known, we can perform a country-specific analysis, and in particular how \textit{close} a country $c$ is to a given item $x$ given its AI investments. This will be given by the normalized sum of the influence of the active AI sectors that are connected to the target $x$. In formula, this corresponds to compute the density \cite{hidalgo2007product} of country $c$'s AI investments around the target item $x$
\begin{equation}
\label{density}
D_{c,x}=\frac{\sum_{x'}M_{c,x'} B_{x',x}}{\sum_{x'} B_{x',x}}
\end{equation}
where we removed the time dependency since we computed this quantity only for the last year of the database. The density values inform about the feasibility of a new trade sector, or the possibility to maintain an already present comparative advantage, given the country's present investments in AI. The more a country invests in sectors $a$ which are strongly connected with the target $x$, the more its AI specialization supports economic development in that sector.
\newpage

\section{Global Results}
\label{sec:global}
\subsection{Pathways for Discovering Specialization in Goods and Services from AI Specialization}
\begin{figure}
[ht]
    \centering
    \includegraphics[height=0.35\paperheight]{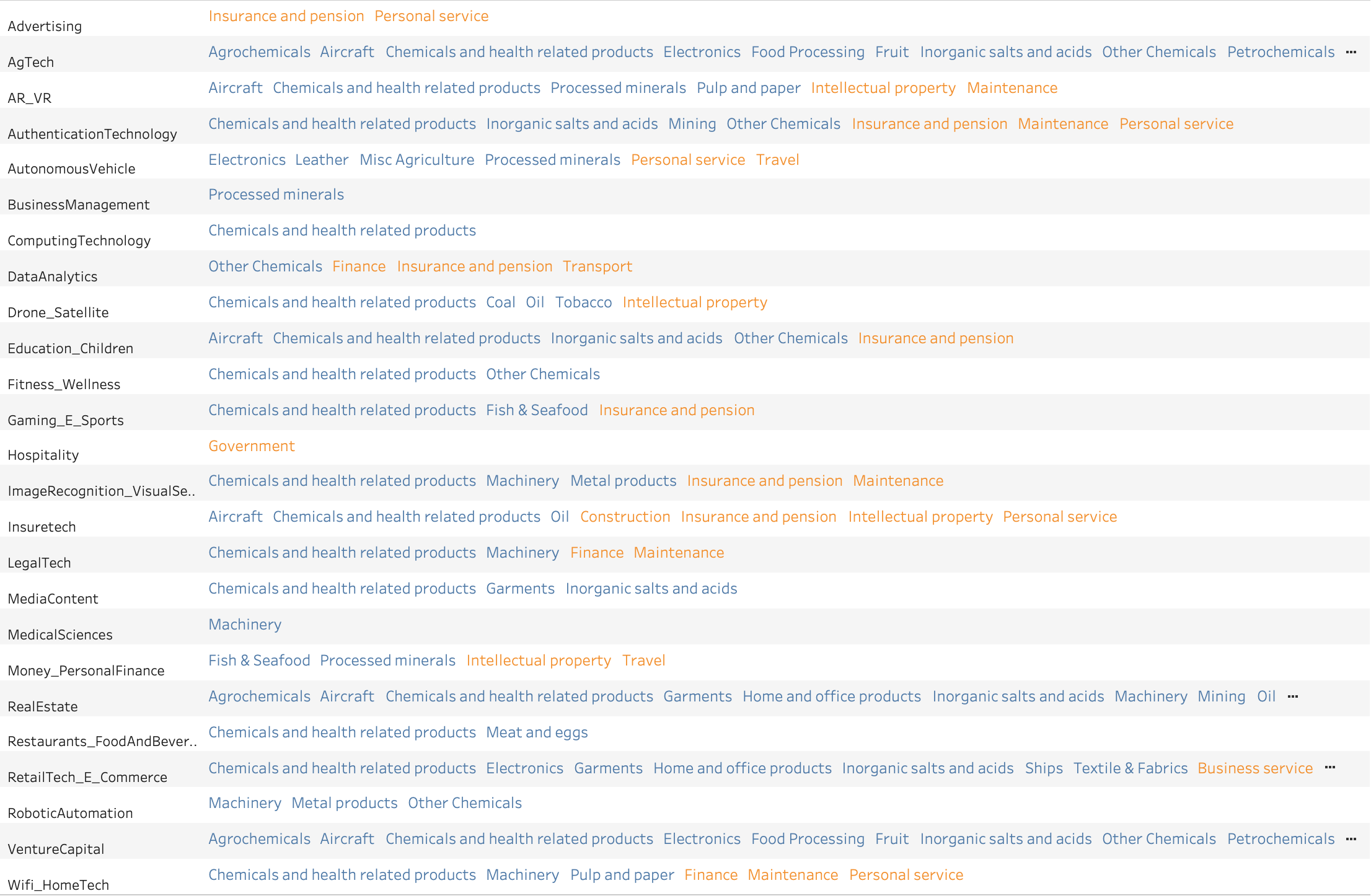}\newline
    \caption{Global co-occurrence from AI investment to goods and service specialization. Column 1 is AI investment, column 2 is associated goods (blue) and services (orange) specialization that are closely linked to the given AI sub-sector of investment. The interpretation is specific and depends on each link. Some links are intuitive to interpret, for example, when countries invest in robotic automation, they subsequently develop specialization in Machinery or Metal products. Interpreting other links could be more subtle, but provides a statistically valid data-driven linkage that requires domain-knowledge in a country-specific context.}
    \label{fig:bipartite}
\end{figure}

\begin{figure}
[ht]
    \centering
    \includegraphics[height=0.7\paperheight]{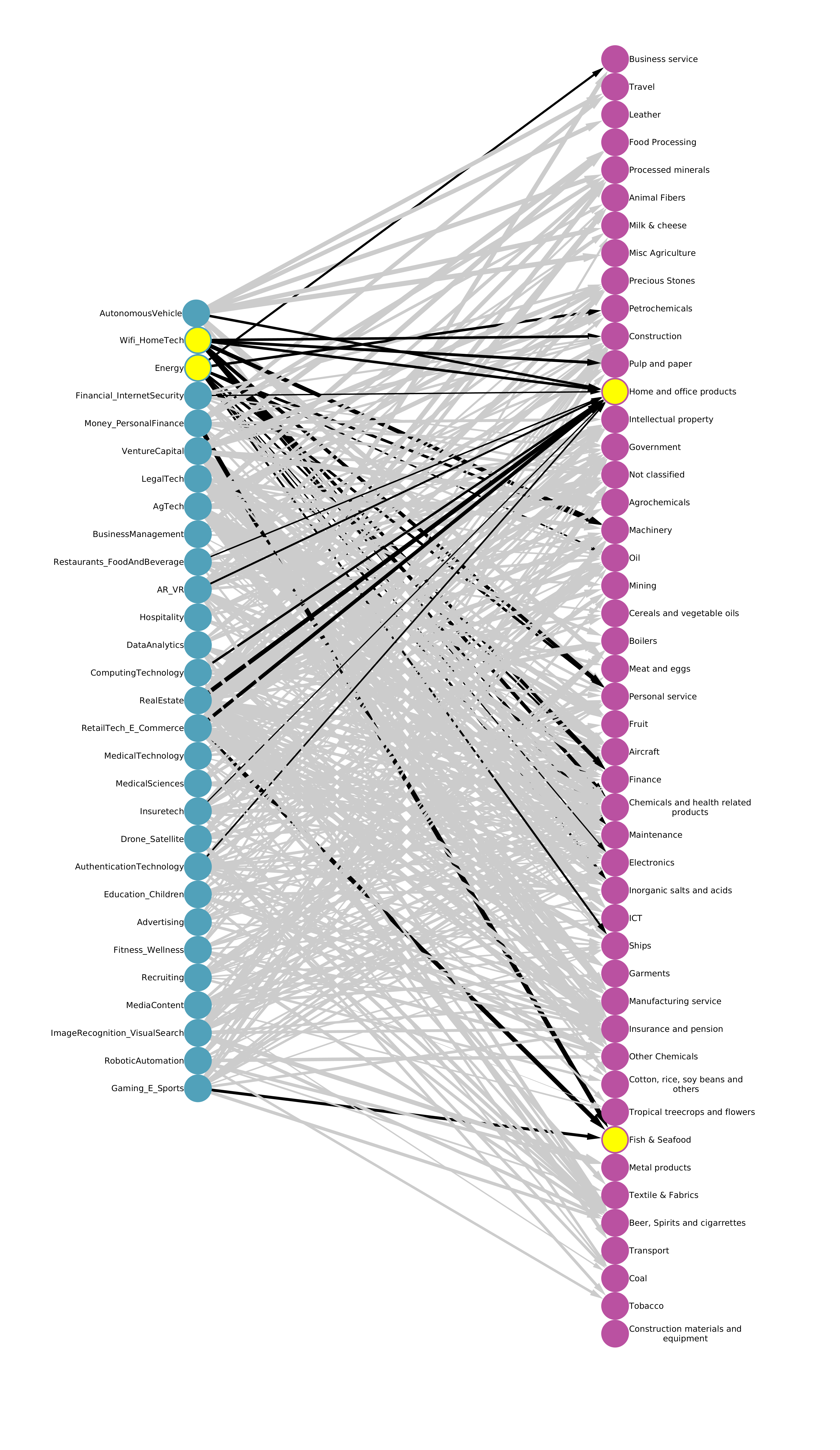}\newline
    \caption{The bipartite network connecting AI (on the left) to Goods and Service (on the right) specialization. Each link has been statistically validated using the Bipartite Configuration Model. For clarity purposes we have highlighted some links and the respective nodes, which we isolate and show in the subsequent figures.}
    \label{fig:global}
\end{figure}

The methodology discssed in the previous section allows us to obtain useful qualitative and quantitative insights on the complex dynamics of AI and broad-based economic specialization. By linking together AI and trade specializations which are related at a given level of statistical significance, we can build the whole network space in which AI, goods, and services are embedded (Figure~\ref{fig:global}). The network shows the linkages from AI investment specialization to goods and service exports specialization. The results imply that when countries invest in AI they subsequently develop capabilities in other sectors (both in other AI sectors and broader economic sectors). The bipartite network presented in Figure \ref{fig:global} shows the most likely connections leading from specialization in AI sub-sectors to  goods and services. Here the width of the links is proportional to the number of times they have been statistically validated. The complex network graph is presented in table \ref{fig:bipartite} format to help discern the single connections in a simpler visualization.

As detailed in the Methods section, two sectors are connected if they are shared by  many countries that are specialized in a given two sectors and follow similar time-based progress in common specialization patterns. From the fact that two sectors are concurrently invested in, or specialization is evident across many countries, we can infer that there is a set of capabilities in common to develop and/or countries develop specialization in similar specializations. After computing the normalized co-occurrences of the Assist Matrix, we statistically validate the links using our null model in order to obtain a relatively sparse network. In the instance of a fully connected network, where all items are connected to each other and time evolution is not explicitly taken into account \cite{hidalgo2007product}, a fully connected network could hinder the understanding of which sector is close to which sector to offer pathways for development. Thanks to the use of the Bipartite Configuration Model, we can use the p-value thresholding to make the network less dense. Finally, our network is also visually interpretable with a p-value threshold that turns to be perfectly reasonable (95\% confidence level). This method offers the advantage to use conditional probabilities that are statistically valid to reliably infer connections such as, on average countries that invest in AgTech, subsequently gain specialization in Agrochemicals, Chemical and health products, Food processing, Fruit, Inorganic salts and acids, Petrochecmicals (Goods); Finance, ICT, or Insurance services (Services). Similarly, countries that invest in Drone and Satellite related AI investment subsequently also gain specialization in Chemical and health products, Coal, Oil (Goods) or Intellectual Property (Services). The interpretability of our approach provides a key tool for policymakers, as we can validate if and how strongly linked each sector is to the other sectors.

\clearpage

\subsection{Within AI Specialization Pathways}

\begin{figure}
[ht]
    \centering
    \includegraphics[height=0.4\paperheight]{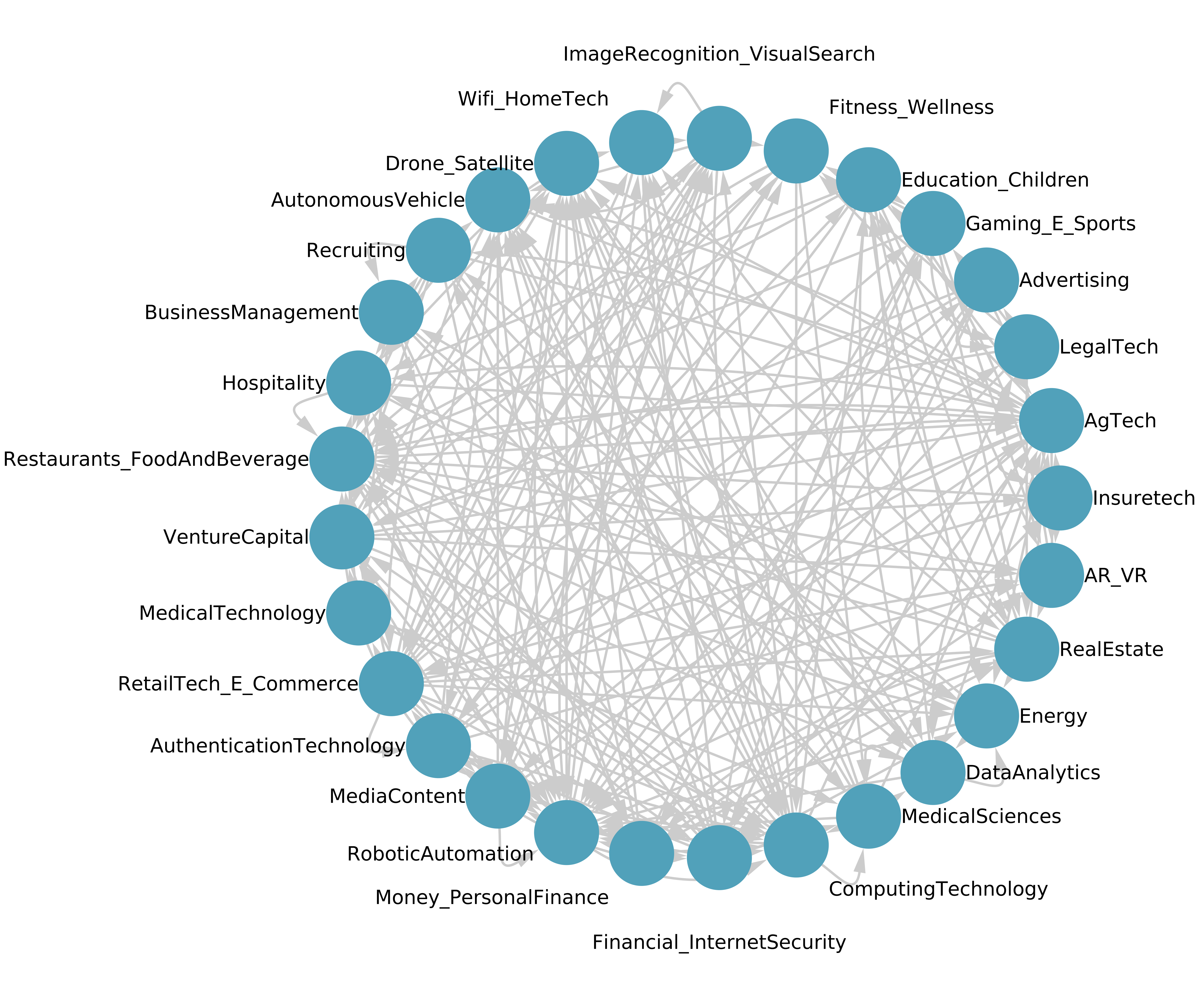}\newline
    \caption{AI based progression network (with time delay). The high number of validated links implies that strong connections exist among most AI sectors i.e., investing in one AI sub-sector usually lead to a subsequent specialization in other AI sub-sectors.}
    \label{fig:aidelay}
\end{figure}

\begin{figure}
[ht]
    \centering
     \includegraphics[height=0.5\paperheight]{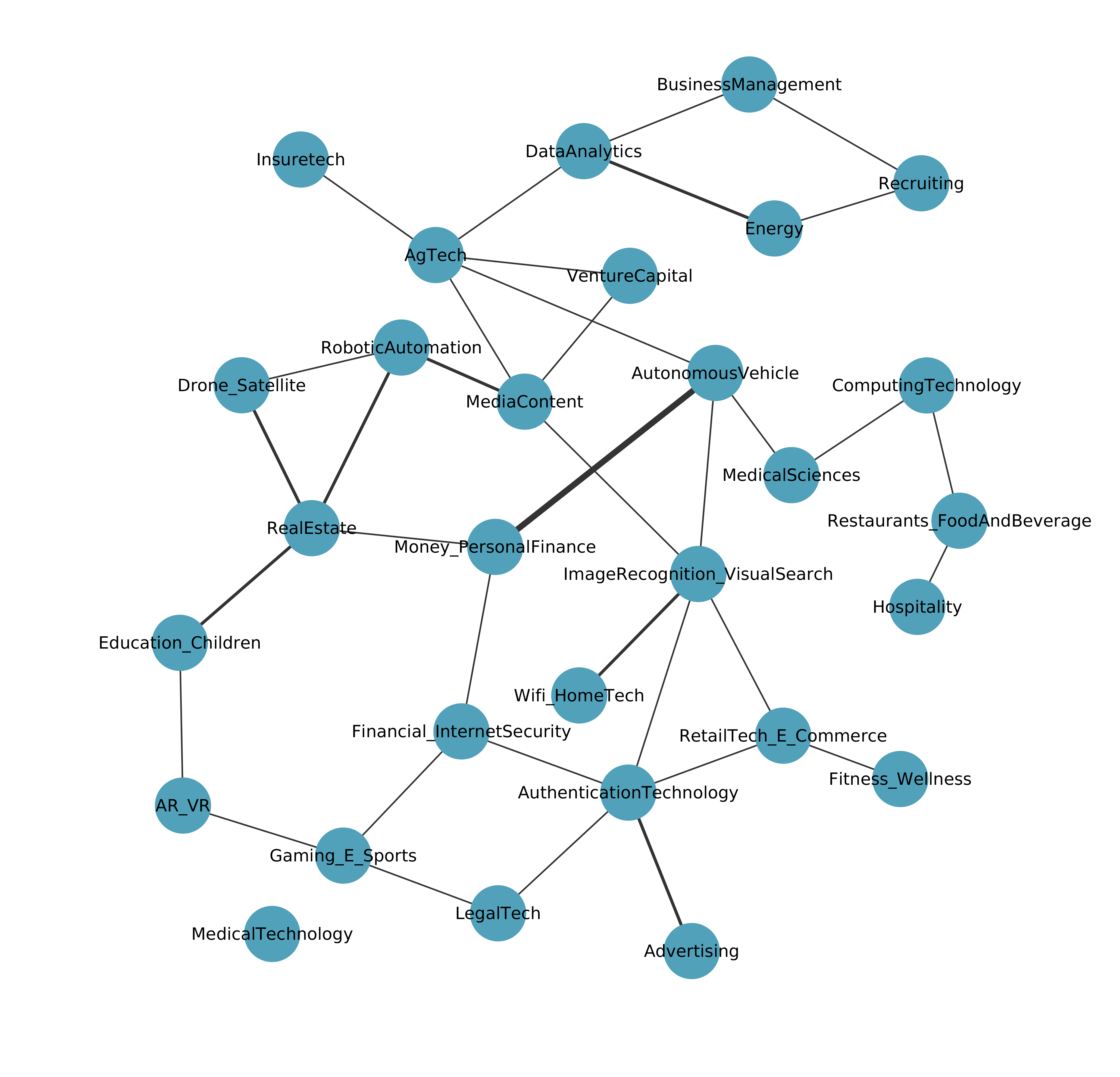}\newline
    \caption{Network of statistically validated co-occurrence in AI specialization, without time delay ($\Delta=0$), so no arrow is present. In this way we can extract the simultaneous co-occurrences and better visualize the central nodes, which could inform a starting point for diversification within AI activities.}
    \label{fig:ainodelay}
\end{figure}

This sub-section presents more focused analysis to show the inter-relation within AI specializations i.e. investment within a specific AI sector leads to gaining specialization in other nearby AI specializations. Figure~\ref{fig:aidelay} shows that many AI sectors are connected to each other. This implies that when a country specializes (or starts investing) in a given AI sub-field, they subsequently also develop specialization in nearby AI capabilities (Figure~\ref{fig:aidelay}). Figure~\ref{fig:aidelay} provides a dynamic progression network which informs the sequence of capabilities learnt by studying the pattern of specialization over time. Note the network has directed arcs that implies sequential progress. For example, specialization in media content related AI subsequently leads to AI related investment in Internet security, or countries that initially invested in Wi-Fi home tech related AI, subsequently started investments in Image Recognition and Visual Search. The directional arrow provides a taste of causal links and can be more clearly interpreted when we look at adjacent nodes, for example, Computing Technology related AI investment can subsequently lead to Medical Science, Data Analytics to Energy related AI investments, or Image recognition to Wi-Fi home tech, Recruiting to Business Management, Hospitality related AI investment leads to Restaurant, Food, and Beverage related AI investments. Figure~\ref{fig:ainodelay} shows the network of statistically validated co-occurrence without the time delay; as a consequence, the links are not directed. This figure provides insights about the capabilities connection or nearby AI sub-sectors that require similar capabilities and can be viewed as the backbone interaction described above. The width of the links is proportional to the number of times the corresponding element of the adjacency matrix has been statistically validated. One can notice that some nodes, such as Hospitality and Recruiting, are more peripheral, meaning that these sectors have a lower degree of interaction with the other sectors, while other nodes, like Money and Personal Finance and Image Recognition and Visual Search, are more central, implying that investing in these sectors allows an easier subsequent diversification in many other AI sectors. To gain diverse portfolio of AI specializations for small-developing countries, the results would imply to begin by investing in one of these central nodes presented in Figure~\ref{fig:ainodelay}.

\newpage
\subsection{Sector Specific Pathways}
\begin{figure}
[h]
\centering
\includegraphics[height=0.22\paperheight]{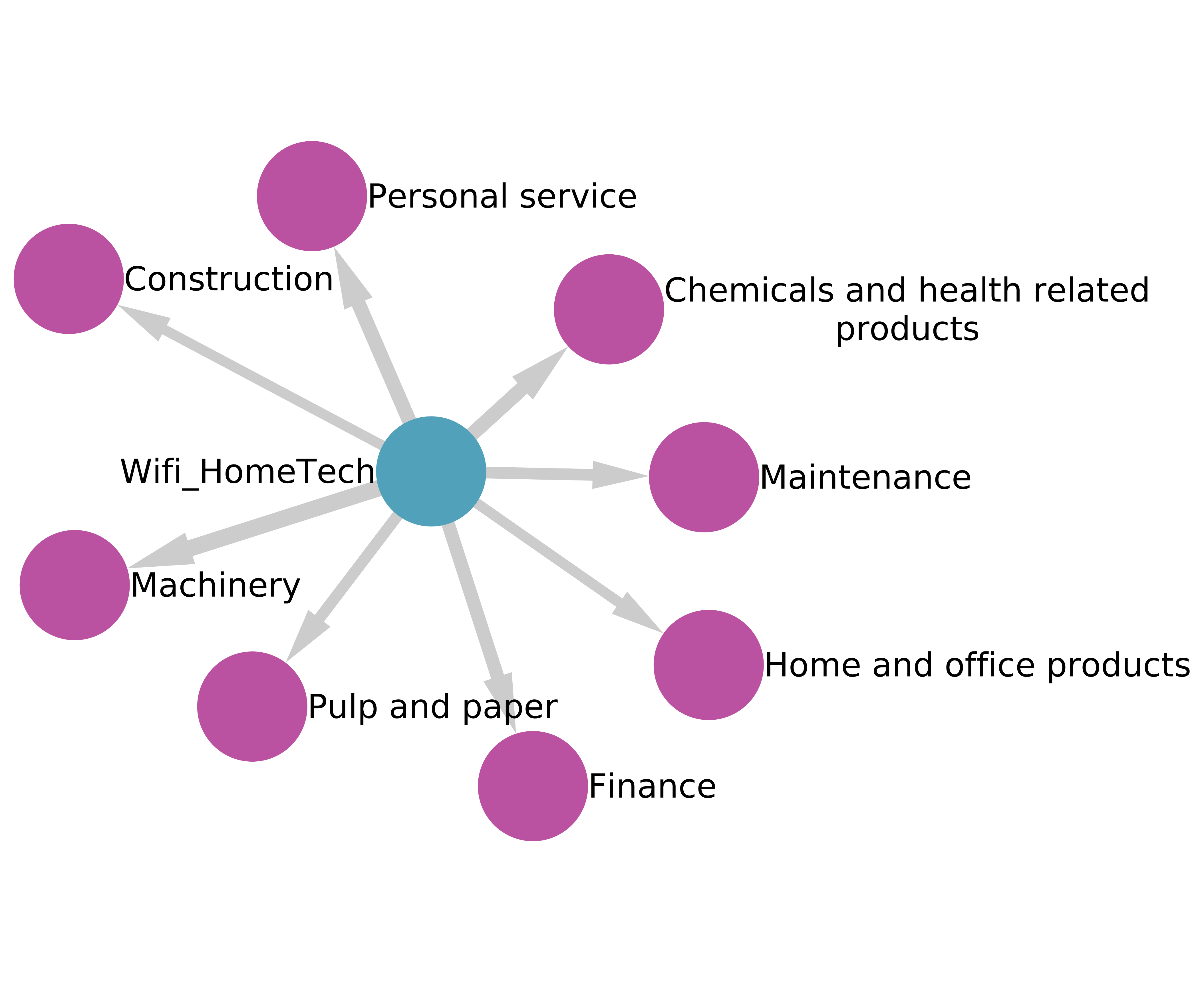}\newline
\includegraphics[height=0.19\paperheight]{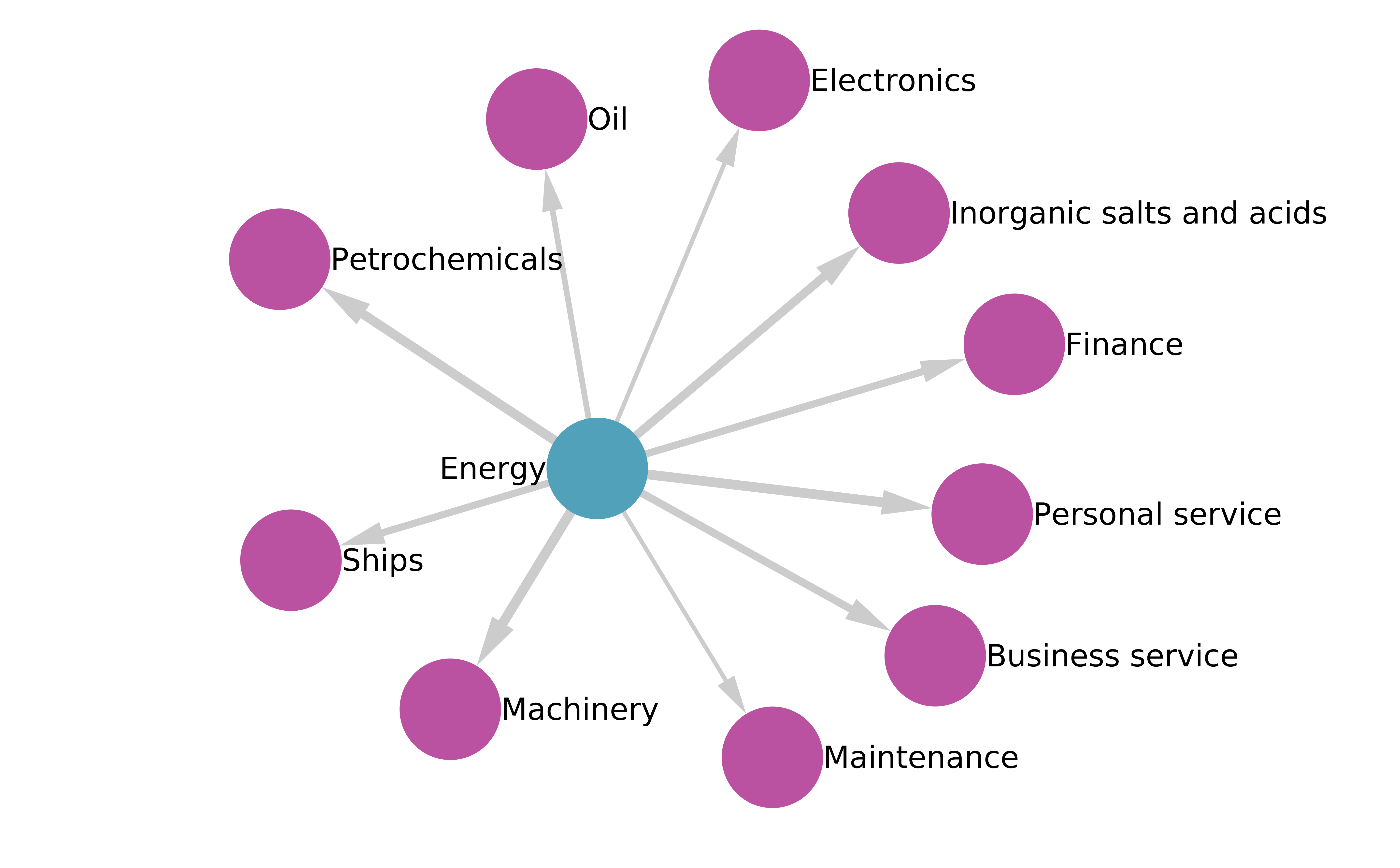}\newline
\includegraphics[height=0.19\paperheight]{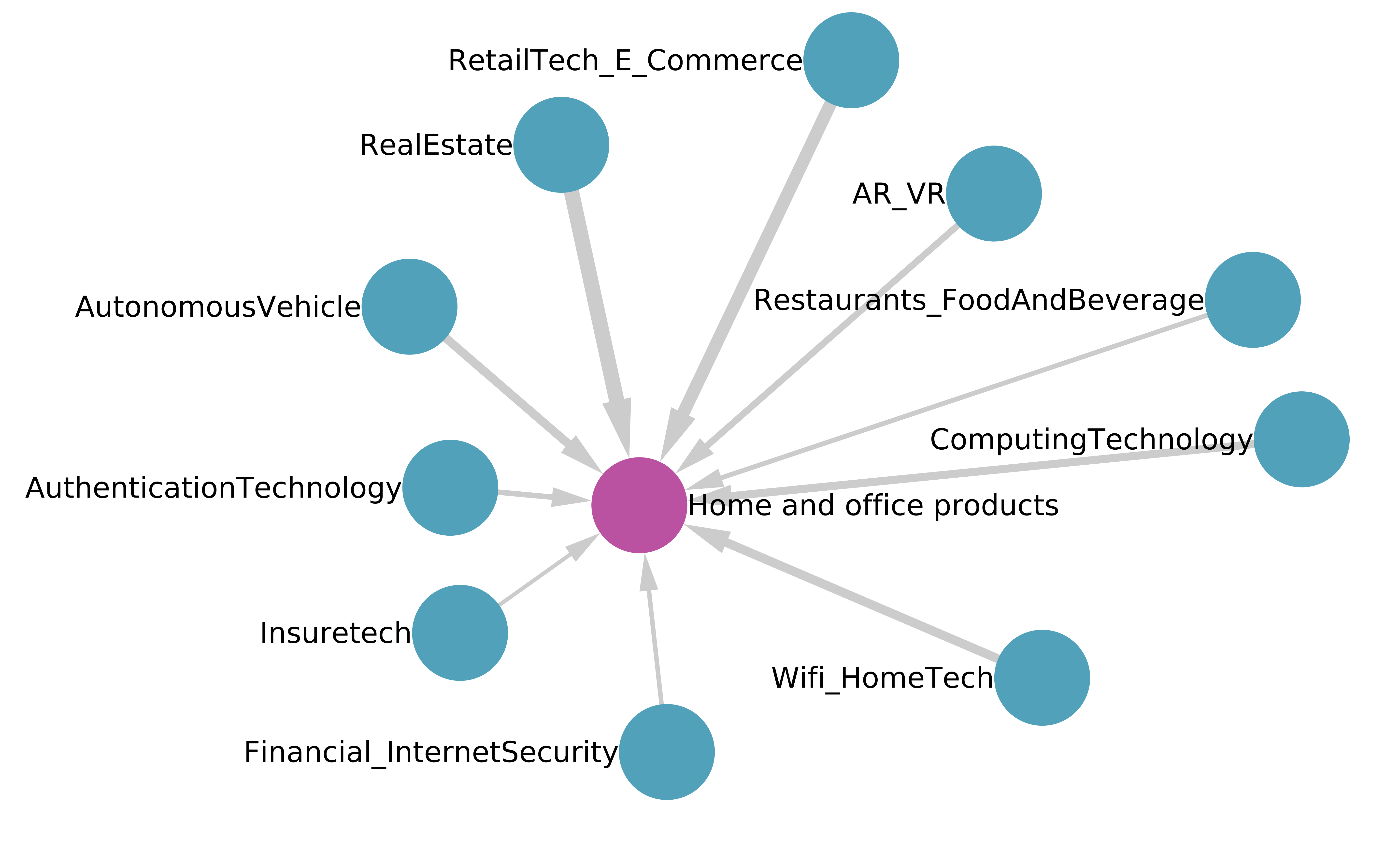}\newline
\caption{The linkage of Wi-Fi Home tech related AI investments to Goods and Services specialization, with time delay (top). The linkage of Energy related AI investments to Goods and Services specialization, with time delay (middle). The linkage of AI investments leading to Home and office product manufacturing  specialization, with time delay (bottom).}
\label{fig:sectors}
\end{figure}

This section presents how the global approach can be broken down into sector specific analyses.
Figure \ref{fig:sectors} provides a zoomed in view of the sequence of specializations across all sectors learnt from the global network of specialization patterns. The top figure shows how specialization in Wifi and Hometech related AI investment subsequently leads to specialization in various sets of goods and services, including Construction, Machinery, Pulp and paper, Chemical and health related products (G) as well as Maintenance, Finance, or Travel related services (S). The middle graph shows the subsequent set of specialization that stem from specialization in Energy related AI investments. Energy related AI investments could lead to gaining specialization in Ships, Petrochemicals, Electronics, Finance, Business, or Maintenance services. Finally, the bottom figure presents an alternative view i.e., that the connection of AI specializations such as Computing Technology, AR/VR, Real Estate, Autonomous Vehicles, Authentication Technology, etc. subsequently lead to specialization in home and office product related manufacturing. Here the width of the link is proportional to the corresponding value of the matrix $\mathbf{B}$.

By learning how strongly interconnected sectors are to each other, we find that once a country starts investing in one sector, the strategy is not to only invest in one sector but develop capabilities many other sectors. Moreover, knowledge and know-hows that a country learns by specializing in one sector can be applied to other sectors. These capabilities become the backbone that offer a unique pathway for countries to develop a portfolio of specializations. Based on the sequence of specialization over time, new links appear, the consequences of specializing in one sector spreads across other sectors of the economy. 

\newpage

\clearpage
\newpage

\section{Country Case Studies}
\label{sec:country}

This section provides an operational framework with country-specific case studies to help design diversification strategies. The cases are presented by stages of economic development. Beginning with an advanced economy, we show the case of Italy, followed by an upper-middle income country --- Mexico, lower-middle income countries --- India and Nigeria. 

\begin{figure}
[ht]
    \centering
    \includegraphics[height=0.20\paperheight]{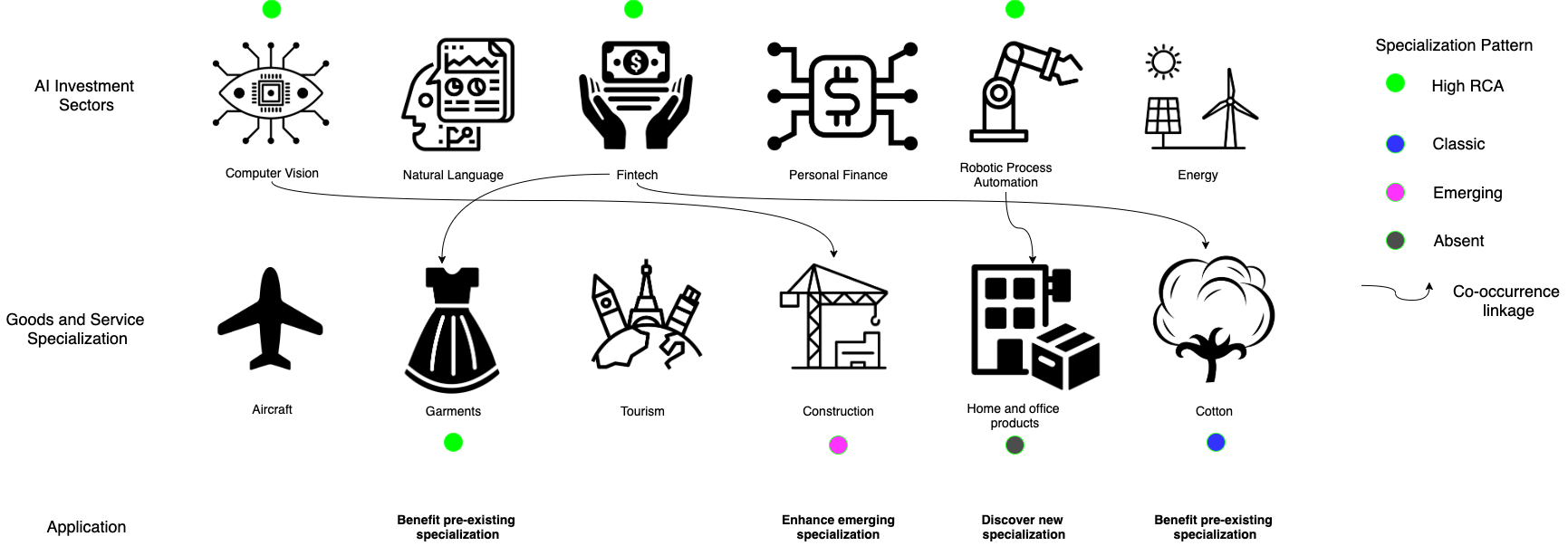}\newline
    \caption{Operational Framework to leverage co-occurrence between specialization in AI, Goods, and Services.}
    \label{fig:opframework}
\end{figure}

Figure~\ref{fig:opframework} shows an abstraction of a country's specialization pattern. The AI investment specialization for varied sub-sectors are presented on the top row where the green dot highlights an active AI specialization i.e. the RCA of private investment in a given AI sub-sector is greater than 1. Example of various goods and service exports specialization are presented in the bottom row. Here, the color codes signify classic, emerging, disappearing, or absent specialization (for goods and services). The links between AI and goods/services signify that the normalized co-occurrence that has been statistically validated based on the global specialization network. The density metric is used to aggregate our findings into a country specific recommendations to highlight the top goods and services, as they relate to the portfolio of AI investment in that country. For example, if there is a strong linkage from fintech to garments and if garment is a classic specialization then these two sectors have strong synergies and this linkage could benefit pre-existing specialization in that country. Similarly, there may be a linkage from fintech to construction services, however, as denoted by the pink dot, the construction industry is an emerging specialization. In this instance, the linkage from fintech to construction may enhance this emerging specialization. The link from robotic process automation to home and office products is valid. However, home and office products are either an absent or disappearing specialization in this hypothetical example. Therefore, the country can leverage its specialization in robotic process automation (RPA) to discover a new specialization in home and office products given it's already specialized in RPA and there exists a link from RPA to home and office products. The density metric is a country-item specific measure.  We can rank the top products and services most closely linked to AI specialization to help answer which sub-AI field can help either (a) enhance pre-existing specialization, or (b) discover a new specialization. 

The country results are presented in two standardized tables. The first country table presents a high-level view of top specializations in a country by ranking the RCA of items (or activities) within AI, Goods, and Services. The RCA measure is based on Ricardian trade theory to reveal the relative difference in productivity. The RCA measure is a first approximation of competitiveness or export (or investment) strength. The RCA helps to avoid the trivial bias which pure export value (or volume) metrics may display, for example, over-representativeness captured by country size or size of the sector. For example, the market share or exports value/volume of a country like the USA or China is very large because of their country size, or in many resource-rich countries, the dominance of oil sector could overshadow other sectors. The RCA measure is an accepted, tried, and tested trade measure, backed by economic theory with practical implications about relative specialization and avoids the bias caused by the sheer size of a given sector. The second table provides a filtered ranking of co-occurrence in goods and services, as they relate to the AI portfolio of the country accounting for the nature of specialization (i.e., Classic, Absent, Emerging, Disappearing). 

\newpage

\subsection{Italy}

\begin{figure}
[ht]
    \centering
    \includegraphics[height=0.25\paperheight]{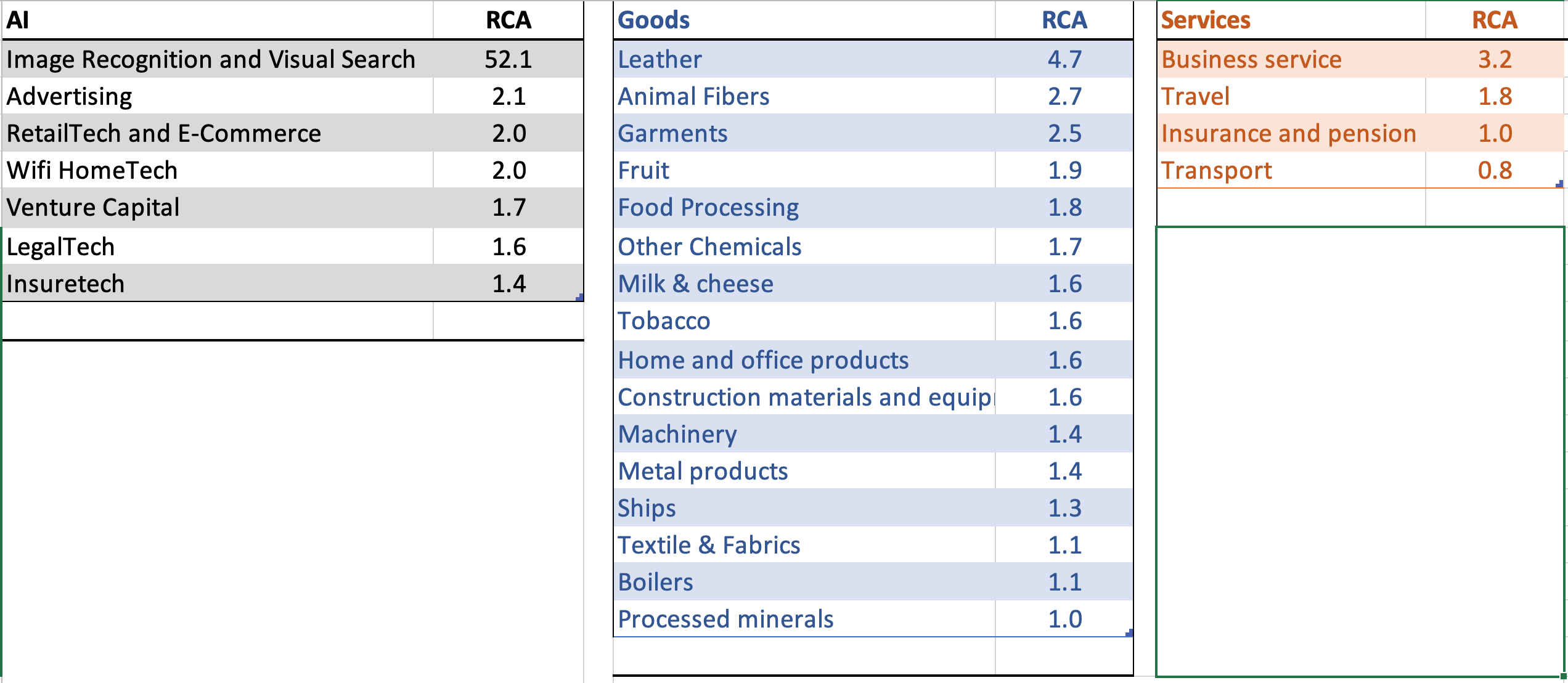}\newline
    \caption{Top areas of specialization in AI, Goods, and Services, Italy, 2019.}
    \label{fig:italy1}
\end{figure}

Figure \ref{fig:italy1} shows the top specialization across sectors in Italy. Italy has developed AI-related investment specialization in Image, Recognition and Visual search, followed by Advertising, Retail tech, e-commerce, and Wifi Home-tech related AI investments. In goods, exports of leather, garments, chemicals, office products, and agricultural production of fruit, food processing, milk and cheese have have high RCA. In services, business services, followed by travel, and insurance services are the top areas of specialization. 

\begin{figure}
[ht]
    \centering
    \includegraphics[height=0.10\paperheight]{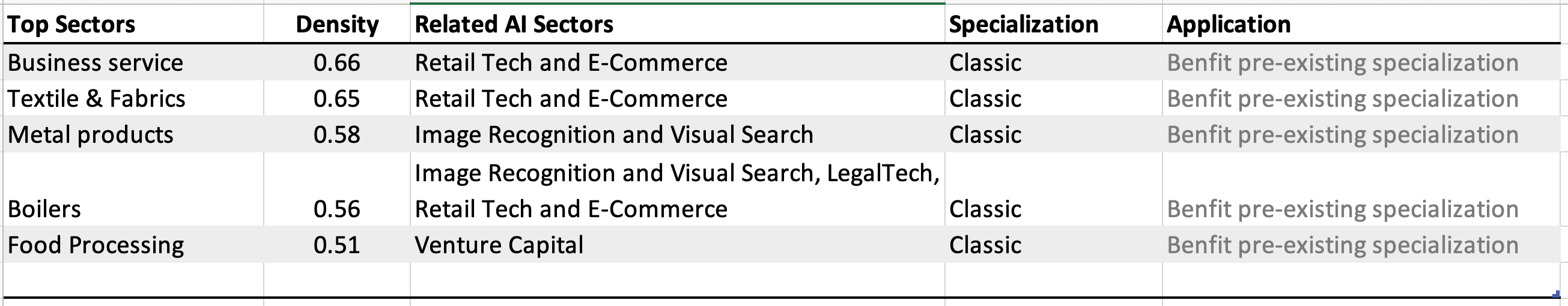}\newline
    \caption{Strongest linkages from portfolio of AI specialization to Goods and Services in Italy. \small {Top sectors (column 1) presents the goods and services most related to the portfolio of AI specializations in Italy based on density measure (column 2) i.e. the feasibility (or likelihood of specialization) given basket of AI specialization. Related AI sectors (column 3) are the adjacent AI sub-sectors related to top sectors presented in column 1. Specialization (column 4) is the current specialization trend of top goods and services sectors presented in column 1; Application (Column 5) is the  recommendation based on linkage between AI, Top Sectors, and Specialization status.}}
    \label{fig:italy2}
\end{figure}

Figure \ref{fig:italy2} presents the strongest linkages from Italy's AI investment portfolio to goods and services. Many links highlight how AI sectors could help Italy retain a competitive edge in sectors of pre-existing specialization. For example, retail and e-commerce can aid business services, textile and fabric industry retain global competitiveness. The deep specialization in image recognition could help metal products, boilers industry, and Venture Capital and fin-tech related AI specialization could aid the food processing industries maintain global competitive edge. 

\newpage
\subsection{Mexico}

\begin{figure}
[h]
\centering
\includegraphics[height=0.13\paperheight]{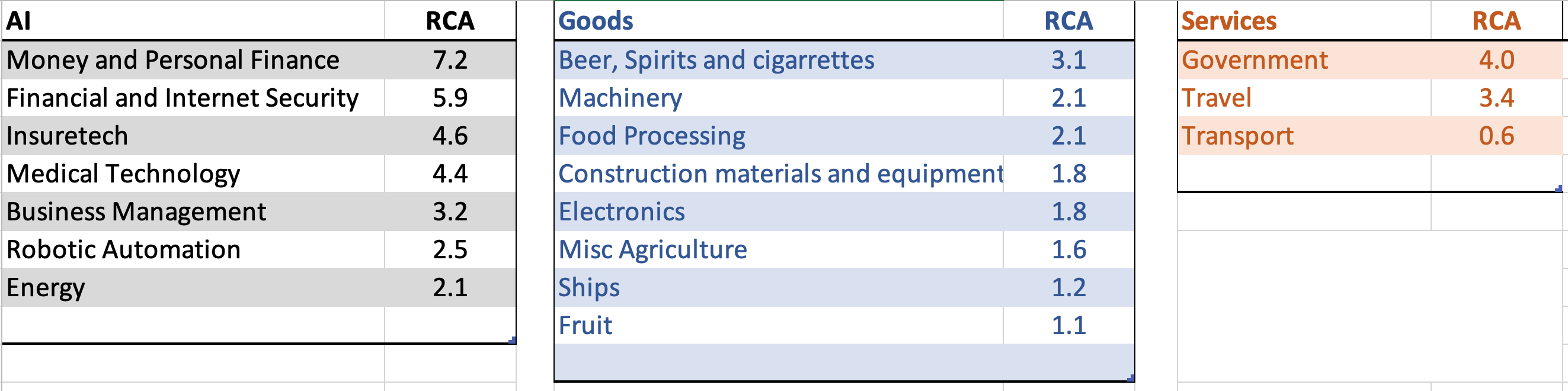}\newline
\caption{Top areas of specialization in AI, Goods, and Services, Mexico, 2019.}
\label{fig:mexico}
\end{figure}

Right mix of industrial policies could help countries like Mexico break out of a potential "middle income trap" (MIT) \cite{arias2015trapped, agenor2012avoiding, felipe2012tracking}.\footnote{MIT is a development stage that characterizes countries that are squeezed between low-wage producers and highly skilled and fast-moving innovators. Caught between these two groups, many middle-income countries are without a viable high-growth strategy and transition becomes difficult \cite{flaaen2013avoid}}. Figure \ref{fig:mexico} shows the top areas of specializations across sectors in Mexico. Mexico has high RCA in many AI sectors, including Money and Personal Finance, Financial and Internet Security, Insure-tech, Medical Technology. In goods, sectors such as Beer, Spirits, and cigarettes, followed by Machinery, Food processing, and Construction material. Oil is the largest classic export in Mexico.\footnote{In the early 1980s, products within the oil community accounted for more than 30 percent of Mexico’s merchandise trade. More recent period, 2015 to 2019, the oil community of products represent only 4.9\% of total merchandise exports. Many AI-related activities could help the oil sector become more competitive.} In services, traditional services such as travel and transport are most dynamic sectors.

\begin{figure}
[h]
    \centering
    \includegraphics[height=0.11\paperheight]{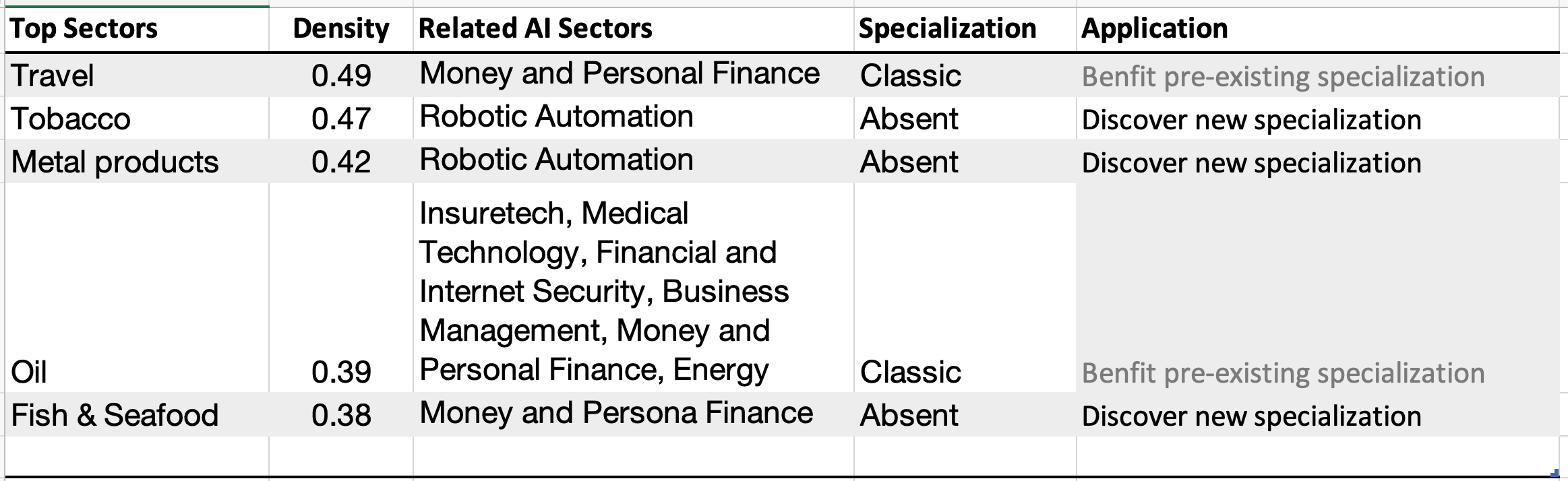}\newline
     \caption{Strongest linkages from portfolio of AI specialization to Goods and Services in Mexico. \small {Top sectors (column 1) presents the goods and services most related to the portfolio of AI specializations in Mexico based on density measure (column 2) i.e. the feasibility (or likelihood of specialization) given basket of AI specialization. Related AI sectors (column 3) are the adjacent AI sub-sectors related to top sectors presented in column 1. Specialization (column 4) is the current specialization trend of top goods and services sectors presented in column 1; Application (Column 5) is the  recommendation based on linkage between AI, Top Sectors, and Specialization status.}}
    \label{fig:mexico2}
\end{figure}

Figure \ref{fig:mexico2} presents the linkage of Mexico's unique AI portfolio that could help classic industries become more competitive and discover new niches. Mexico's unique AI capabilities including in fin-tech, insure-tech, internet security, medical technology, energy etc. could help the travel services and ecosystem related to the oil industry remain competitive, whereas, robotic automation related investment specialization could aid to discover new (or disappearing) specializations such as in metals, or fin-tech related specialization help re-gain specialization in fish and seafood industry.\footnote{The number of fin-tech and insure-tech startups have been booming in Mexico. Fin-tech startups include in sub-sectors of Payments and Remittances, Lending, Enterprise Financial Management, Personal Financial Management, Crowdfunding, etc.} Garnering the diversity of AI investments in Mexico could help avoid the "middle-income trap" by formulating pathways for broad based economic diversification.

\newpage
\subsection{India}

\begin{figure}
[h]
    \centering
    \includegraphics[height=0.25\paperheight]{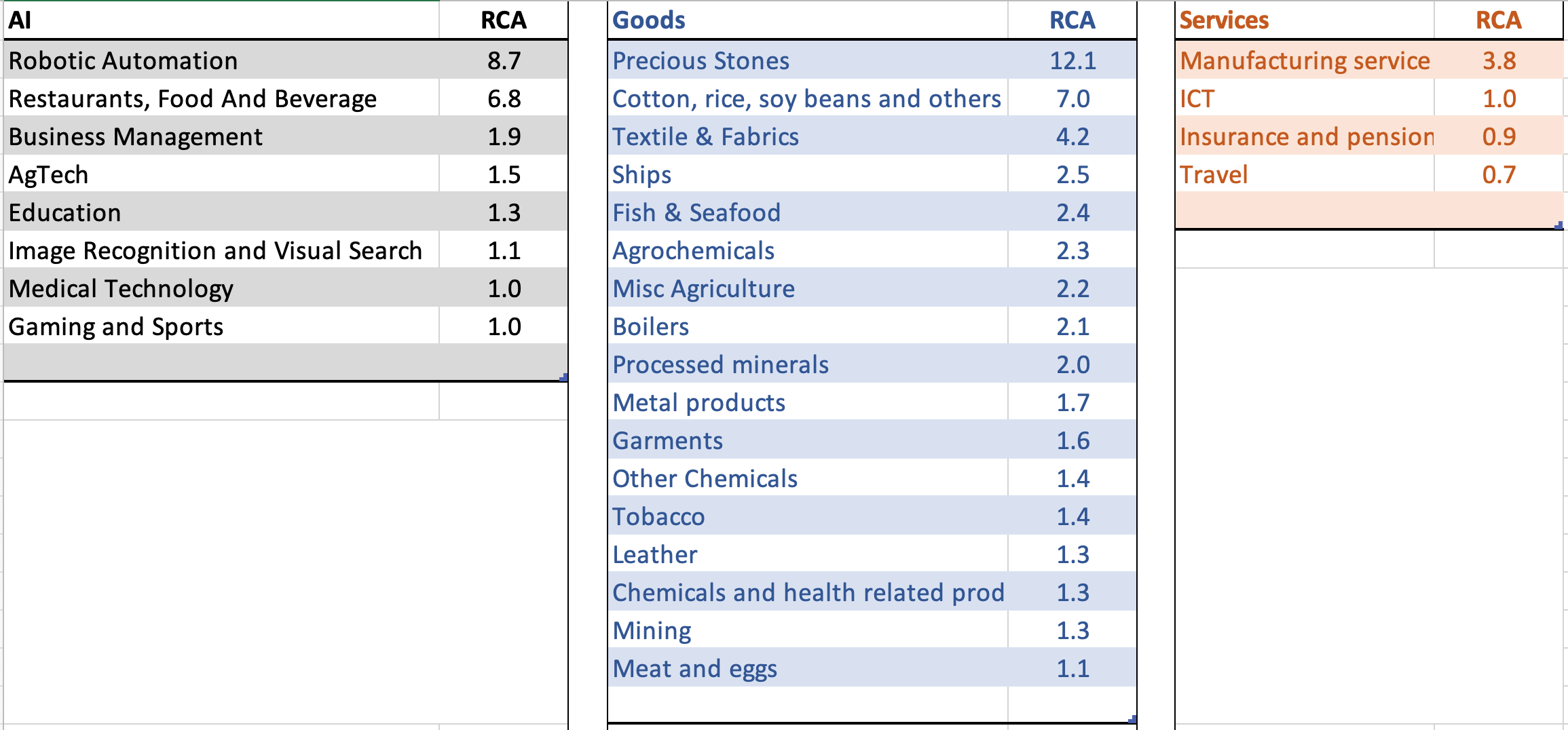}\newline
    \caption{Top areas of specializations in AI, Goods, and Services, India, 2019.}
    \label{fig:india1}
\end{figure}

India has been progressively diversifying its goods and service exports \cite{anand2015make}. Peculiar to India's specialization pattern is the size and sophistication of it's service exports \cite{mishra2011service}. Figure \ref{fig:india1} shows the top specializations for India across sectors. Robotic Automation, Restaurants, Food, and Beverage, Business Management, and AgTech related AI investments are most prominent. In goods, primary resource-based and low-medium tech manufacturing sectors are eminent, including precious stones for jewellery, cotton based products, textile and fabrics, agrochemicals, etc. Services remain a relatively large portion of exports but are concentrated in manufacturing, computer service, insurance, and tourism \cite{anand2015make, mishra2011service, mishra2020economic}. 

\begin{figure}
[h]
    \centering
    \includegraphics[height=0.15\paperheight]{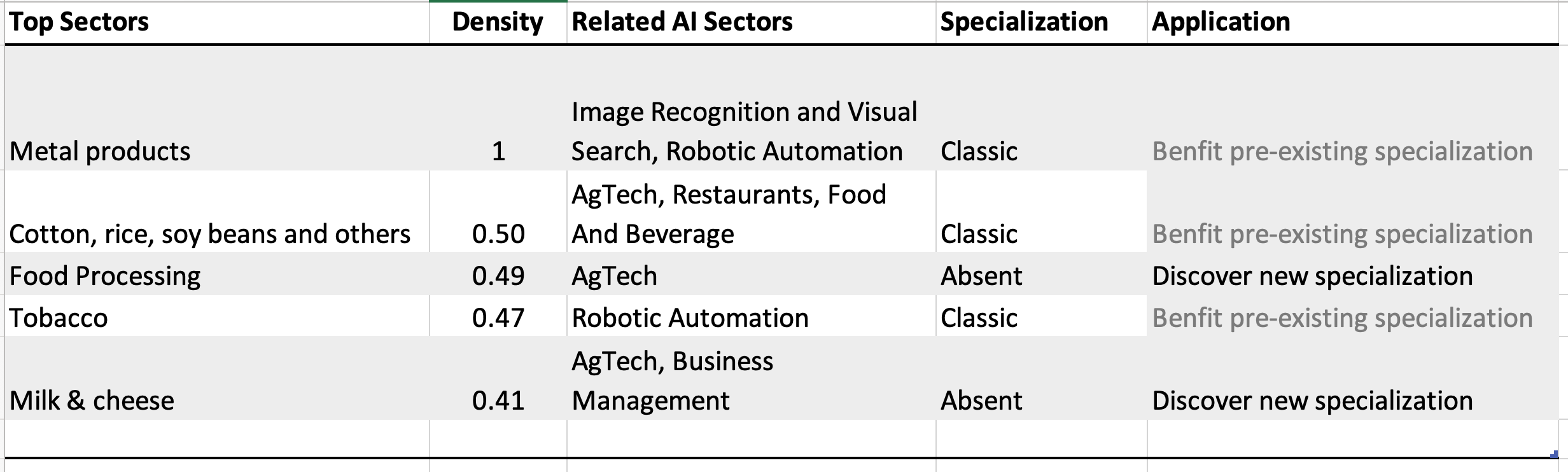}\newline
    \caption{Strongest linkages from portfolio of AI specialization to Goods and Services in India. \small {Top sectors (column 1) presents the goods and services most related to the portfolio of AI specializations in India based on density measure (column 2) i.e. the feasibility (or likelihood of specialization) given basket of AI specialization. Related AI sectors (column 3) are the adjacent AI sub-sectors related to top sectors presented in column 1. Specialization (column 4) is the current specialization trend of top goods and services sectors presented in column 1; Application (Column 5) is the  recommendation based on linkage between AI, Top Sectors, and Specialization status.}}
    \label{fig:india2}
\end{figure}

Figure \ref{fig:india2} presents examples of how India's unique AI capabilities could potentially help diversify and make manufacturing and services specialization more competitive. India's capabilities in Robotic Process Automation related AI investments could help benefit high-tech and medium-tech manufacturing in pre-existing specializations, such as, metal, agro-related industries including cotton, or manufactured tobacco. Specialization in AgTech could help the agriculture and informal sector by discovering new niches in Food processing, Milk and cheese, and related agro and chemical products.  

\newpage
\subsection{Nigeria}

\begin{figure}
[h]
    \centering
    \includegraphics[height=0.15\paperheight]{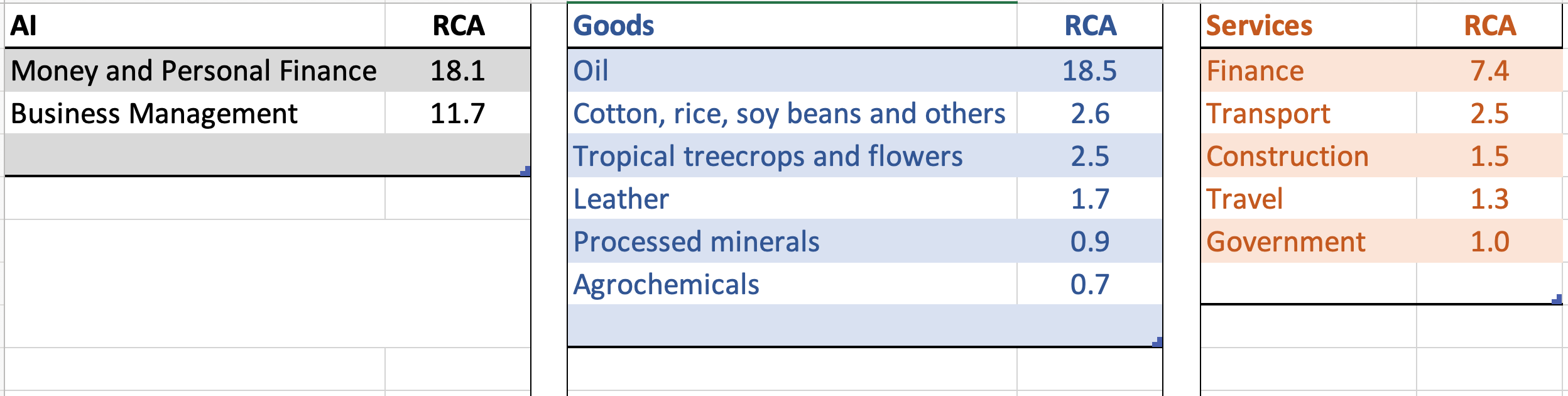}\newline
    \caption{Top areas of specializations in AI, Goods, and Services, Nigeria, 2019.}
    \label{fig:nigeria}
\end{figure}

Resource rich countries like Nigeria (or Mexico) face unique structural transformation challenges \cite{nigeriarticle4}.\footnote{Structural transformation is broadly defined as the reallocation of resources from low to high value added tasks or sectors.  For  recent  discussions  of  the  importance  of  structural  transformation  and  development,  see  \cite{rodrik2013structural, battaile2015transforming, gelb2010economic, timmer2008structural}. Resource sectors tend to be highly capital intensive and offer limited employment opportunities to accommodate workers exiting sectors with lower average productivity, such as agriculture and informal services leading to significant "Dutch disease" effects, stemming from a shift in demand following a resource discovery \cite{battaile2015transforming}.} Figure \ref{fig:nigeria} shows the top specializations in Nigeria. Nigeria shows strong AI investment-related specialization in Money and Personal finance, and Business Management related AI clusters that could help guide pathways for diversification. Sectors such as Oil, Cotton, rice, tropical tree crops, leather sector exhibit high RCA in goods; Finance, Transportation, and Construction in services. 

\begin{figure}
[h]
    \centering
    \includegraphics[height=0.15\paperheight]{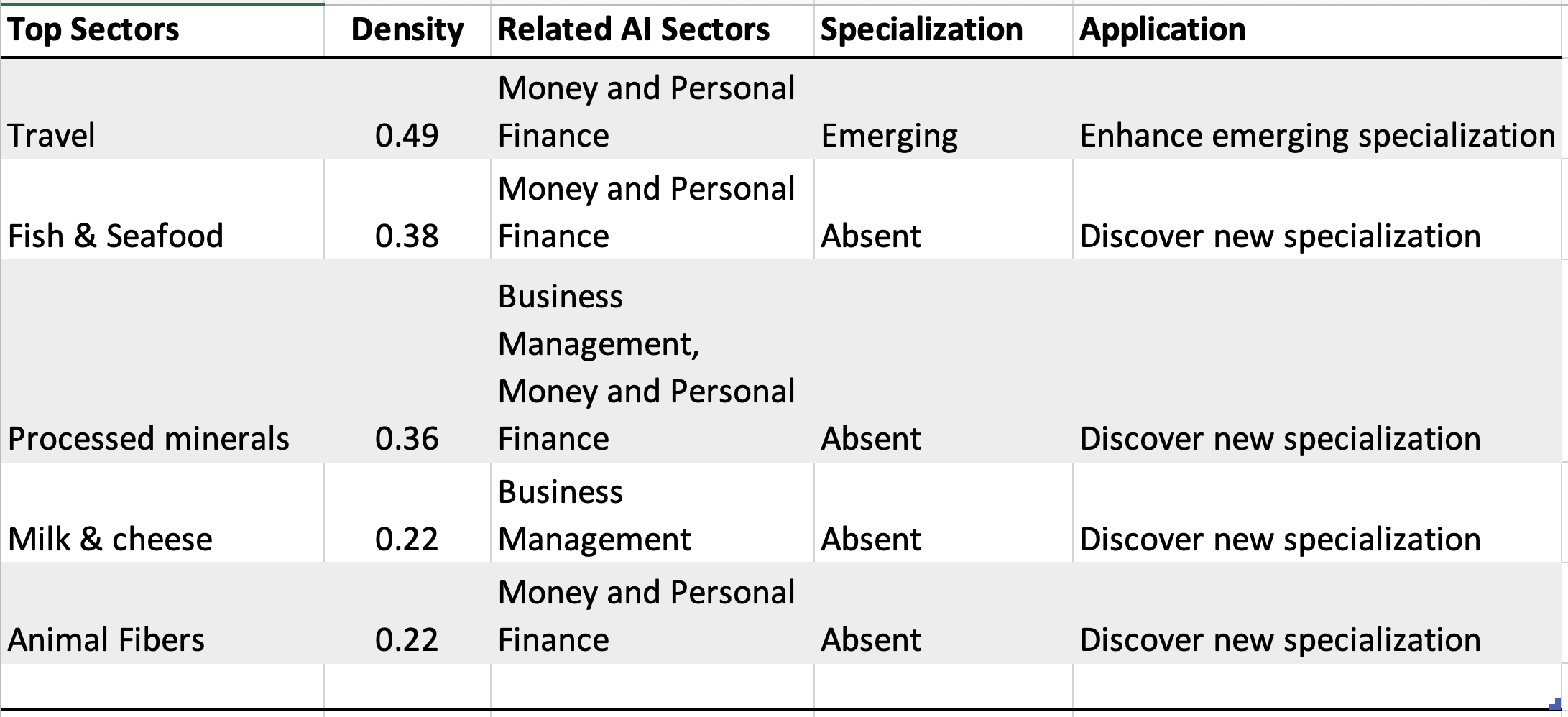}\newline
    \caption{Strongest linkages from portfolio of AI specialization to Goods and Services in Nigeria. \small {Top sectors (column 1) presents the goods and services most related to the portfolio of AI specializations in Nigeria based on density measure (column 2) i.e. the feasibility (or likelihood of specialization) given basket of AI specialization. Related AI sectors (column 3) are the adjacent AI sub-sectors related to top sectors presented in column 1. Specialization (column 4) is the current specialization trend of top goods and services sectors presented in column 1; Application (Column 5) is the  recommendation based on linkage between AI, Top Sectors, and Specialization status.}}
    \label{fig:nigeria2}
\end{figure}

Leveraging capabilities in the oil, agriculture, and infrastructure-related sectors, AI could be strategically important asset to aid competitiveness of pre-existing sectors and discover new ones. Figure \ref{fig:nigeria2} shows the strongest linkages to Nigeria's AI investment portfolio. Travel and tourism related services are emerging specialization where investment in Money and Personal finance could help make tourism-related specialization more resilient. Personal finance and business management related AI investment could also aid discovery in new capabilities. For example, fish and seafood, processed minerals, or animal fibres are linked to initial investment in fin-tech related AI capabilities. Current AI specializations could help planned physical infrastructure investments, and help discover pathways to other AI sectors that in turn could aid broader diversification strategies. 

\newpage

\section{Discussion}
\label{sec:disucssion}

The paper provided a first of a kind novel integration of data and methodology to examine the relationship of AI based specialization to broad based economic specialization. The results have practical implications for potential strategic resource allocation and policy decisions. Technological changes are driving up the demand for AI-based services. AI services can be developed in one location but consumed in many other places. Historically buyers and sellers needed to be face to face. However, AI services between buyers and sellers can be traded globally across and within borders almost instantly. Increasingly, AI services are embedded in various manufacturing and service-based products that are digitally traded. Our study offers hope for developing countries to leverage AI specialization as an opportunity to diversify the sources of comparative advantage. The paper documented  how sector-specific AI investments not only lead to subsequent specialization in other AI investments but could also lead to developing capabilities in goods and services. The paper also offered an operational framework with country case studies to present how inherent AI capabilities could help pre-existing industries become more competitive and/or discover new areas of specialization. The systematic framework can directly aid growth and competitiveness strategies for nations across all stages of development.  

In Silicon Valley, there is a truism that “70 percent of hardware is software” ---  this was the early recognition of the tight link between sales of computers and software services. The life-cycle and adoption of AI at a national-level or in enterprises relies on a ecosystem of services to translate into broad-based economic transformation. AI-services are the becoming glue that binds many manufacturing supply chains and critical to their reliable operations. The future wealth of nations could very well depend on an underpinning of AI components, from research and development at the inception of the product to automated distribution and repair at completion. Our paper combines traditional view of industrial strategies and growing wave of national AI strategies to offer a more complete data-driven view that may be critical to shape the future growth patterns of nations. This research also has implications for the Venture Capital community to help design more macro-focused investment strategies accounting for current AI capabilities and broader economic capabilities to aid strategic diversification and value creation opportunities. 

This work has helped open many areas of future research. This paper focused on within AI specialization and from AI to  broader economic specialization linkages. Future research could explore how goods and services based capabilities can lead to discovery of new AI based specializations. This research focused on AI investment as a proxy for AI specialization but future studies could also incorporate scientific discoveries, i.e., specialization in AI stemming from patents, journal publications, conference papers. Nevertheless, investment data is more robust and relevant as it provides a stronger signal of economic activity that is sensitive to domestic companies (or startups) that are fuelling the wave of AI diffusion within and across countries. Future research could also provide deeper country specific case strategies that will help validate the specific linkages presented in this paper based on industry-specific domain-knowledge and national context. There are unanswered technical challenges. For example, given different data sources and units for AI investment (versus exports data) pose a constraint to combine all the data in a unified matrix. There are also open questions about implications of this study, such as the trade-offs involved for developing countries to import AI capabilities versus producing their own AI domestically. The real-world application areas based on the identified linkages between AI and broader economic sectors are also worthy of future research.



\newpage
\bibliographystyle{unsrt}
\bibliography{references.bib}

\newpage
\appendix

\section*{Technical Appendix}
\label{sec:appendix}
\renewcommand{\thesubsection}{\Alph{subsection}}

\subsection{Data Description}

\subsubsection{AI Investment}

Organization data is embedded from CapIQ and Crunchbase. These companies include all types of companies (private, public, operating, operating as subsidiary, out of business) in the world; The investment data include private investment, Mergers \& Acquisitions (M\&A), public offering, minority stake made by Private Equity/Venture Capital firms, corporate venture arms, governments, and institutions across more than 80 countries. Some data is simply unreachable when the investors are undisclosed, or the funding amounts by investors are undisclosed. We also embed firmographic information such as year founded and HQ location. Quid embeds CapIQ data as a default and we add in data from Crunchbase for the ones that are not captured in CapIQ. This way we not only have comprehensive and accurate data on all global organizations, but also capture early-stage startups and funding events data that are publicly available. 

A boolean search for "artificial intelligence" OR "AI" OR "machine learning" OR "deep learning” yields our source of private investments in AI. Private investment is a private sale of newly issued securities (equity or debt) by a company to a selected investor or a selected group of investors. The stakes that buyers take in private placements are often minority stakes (under 50\%), although it is possible to take control of a company through a private placement as well, in which case the private placement would be a majority stake investment. As AI investment is a relatively new field, our data is focused on the time-period between 2010-2020. We have open-sourced the data for the broader research community. 

\subsubsection{Service Exports}

The source of the service exports analysis relies on Balance of Payments Manual (BPM6) classification of disaggregated data on credit accounts of services at 1-digit level. The BPM6 provides credit, debit and net accounts for services between 2010-2020. The coverage of trade in services data is still very limited and often unbalanced. Services, in contrast to goods, are characterized by several features, such as intangibility and non-storability, which complicate the collection of accurate international trade in services statistics \cite{spinelli2015estimating}. In the spirit of \cite{stojkoski2016impact, zaccaria2018integrating, mishra2020economic} the combined goods and services dataset of world trade has been referred to as the universal matrix of world trade. 

\subsubsection{Goods Exports}

Merchandise trade data was obtained from the U.N. Comtrade database available through the World Bank World Integrated Trade Solution System (WITS\footnote{The World Integrated Trade Solution (WITS) software provides access to international merchandise trade, tariff and non-tariff measures (NTM) data. \url{https://wits.worldbank.org}}). The Standard International Trade Classification (SITC) Revision 2 is preferred, which contains 786 items at its 4-digit level. The SITC Revision 2 was published in 1975 and the majority of countries continue to use it for various purposes, such as study of long-term trends in international merchandise trade and aggregation of traded commodities into classes more suitable for economic analysis. More recent revisions, nonetheless, have new set of product categories that have been created in recent years.\footnote{The SITC classification reflects (a) the materials used in production, (b) the processing stage, (c) market practices and uses of the products, (d) the importance of the commodities in terms of world trade, and (e) technological changes. \url{ https://unstats.un.org/unsd/classifications/Family/Detail/1070 }} Since countries report both their imports and their exports to the United Nations, the raw data duplicates flows . As a result, there is a process for reconciling data due to well-known discrepancies in trade statistics. In general terms, we consider mirror statistics for fragile countries low-income countries.

Using the trade data, standard metrics of trade were constructed following \cite{hausmann2006structural} a series of network metrics were calculation on top of the merchandise trade data, such as revealed comparative advantage, proximity, and density. In addition, the product space shows that products group naturally into highly connected communities. This suggest that product in these communities are closely connected to each other than products outside of the community. Following \cite{hausmann2014atlas} product communities complement the trade database. These communities were originally created using the two-step algorithm introduced by \cite{rosvall2008maps}.

\clearpage
\newpage

\subsubsection{Appendix Tables and Figures}

\begin{longtable}{p{.3\linewidth}p{.7\linewidth}}
\caption{AI sectors used and their sub-sectors aggregation} \\
  
\footnotesize AI Investment sectors used     &     AI sub-sectors included \\
\footnotesize Advertising &\footnotesize Advertisers, Programmatic, Mobile advertising, Real time bidding, Influencer, Social media marketing, Content marketing, Advertisers \\
\footnotesize AgTech &\footnotesize Agriculture, Farmers, Farming, Crop \\
\footnotesize AR, VR &\footnotesize Augmented reality, Vr, Virtual reality, Ar \\
\footnotesize Authentication Technology &\footnotesize Facial, Face recognition, Law enforcement, Video surveillance \\
\footnotesize Autonomous Vehicle &\footnotesize Autonomous vehicles, Fleet, Road, Autonomous driving \\
\footnotesize Cancer, Drug Discovery &\footnotesize Drug, Cancer, Therapy, Genomic \\
\footnotesize Computing Technology &\footnotesize Quantum, Quantum computing technologies, Applications for quantum, Simulation of quantum, Semiconductor, Chips, Processors, Low power \\
\footnotesize Crypto, Wealth Management &\footnotesize Crypto, Wealth management, Traders, Cryptocurrency \\
\footnotesize Data Science, Data Platform &\footnotesize Data centers, Migration, Cloud management, Application performance, Reinforcement learning, General intelligence, Sift through data, Platform for AI, Sql, Hadoop, Python, Data preparation \\
\footnotesize Drone, Satellite &\footnotesize Drone, Satellite, Unmanned, Remote sensing \\
\footnotesize Education &\footnotesize Student, Edtech, Children, Career \\
\footnotesize Energy Management &\footnotesize Energy management, Buildings, Renewable, Electricity \\
\footnotesize Fitness and Wellness &\footnotesize Wellness, Wearable, Fitness, Emotions \\
\footnotesize Fraud Detection, Money Laundering &\footnotesize Fraud detection, Merchants, Laundering, Personal finance \\
\footnotesize Gaming, e-sports &\footnotesize Player, Esports, Mobile games, Fans \\
\footnotesize Hospitality &\footnotesize Hotels, Booking, Business travel, Online travel \\
\footnotesize Insuretech &\footnotesize Insurtech, Insurance industry, Underwriting, Insurance products \\
\footnotesize Legal Tech &\footnotesize Legal, Law, Contract management, Lawyers \\
\footnotesize Media Content &\footnotesize Editing, Instagram, Photo sharing, Reserve a table, Topics, Video content, Readers, Personalized content \\
\footnotesize Medical Technology &\footnotesize Doctors, Hospital, Physicians, Medication, Medical device, Surgical, Blood, Cardiac \\
\footnotesize Money, Personal Finance &\footnotesize Invoices, Medium businesses, Cash flow, Receipts, Lending, Loans, Credit score, Consumer finance \\
\footnotesize Network Security &\footnotesize Threat, Network security, Cybersecurity, Security solutions \\
\footnotesize Oil and Gas &\footnotesize Gas, Predictive maintenance, Industrial automation, Machinery \\
\footnotesize Real Estate &\footnotesize Commercial real estate, Landlords, Estate agents, Property management \\
\footnotesize Recruiting &\footnotesize Recruiting, Candidate, Hiring process, Recruiters \\
\footnotesize Restaurants, Food and Beverage &\footnotesize Food and beverage, Kitchen, Grocery, Food delivery \\
\footnotesize Retail Tech, E-Commerce &\footnotesize Ecommerce, Marketing automation, Shoppers, Retail technology \\
\footnotesize Robotic Automation &\footnotesize Industrial automation, Ai robotics, Mobile robot, Warehouse management, Robotic process automation, Rpa, Test automation, Business process automation \\
\footnotesize Sales Automation, Meeting Management &\footnotesize Meetings, Sales automation, Prospects, Sales teams \\
\footnotesize Supply Chain Management &\footnotesize Supply chain management, Freight, Shipping, Procurement process \\
\footnotesize Text Analytics &\footnotesize Bots, Chatbots, Conversational ai, Messenger, Sentiment, Customer feedback, Employee experience, Text analytics, Speech recognition, Musical, Podcasts, Songs \\
\footnotesize Venture Capital &\footnotesize Equity, Technology startups, Mentorship, Platform for startups \\
\footnotesize Wifi, Home Tech &\footnotesize Requires ios, Ipod touch, Requires ios compatible, Compatible with iphone, Wi fi, Indoor, Wifi, Lights \\
\footnotesize Image Recognition, Visual Search &\footnotesize Fashion, Visual search, Shoes, Apparel \\
\footnotesize Neuroscience &\footnotesize Palo alto, Semantic analysis technology, Knowledge mapping, Public opinion monitoring \\
\label{tab:aiappendix}
\end{longtable}

\begin{figure}
    \centering
    \includegraphics[height=0.35\paperheight]{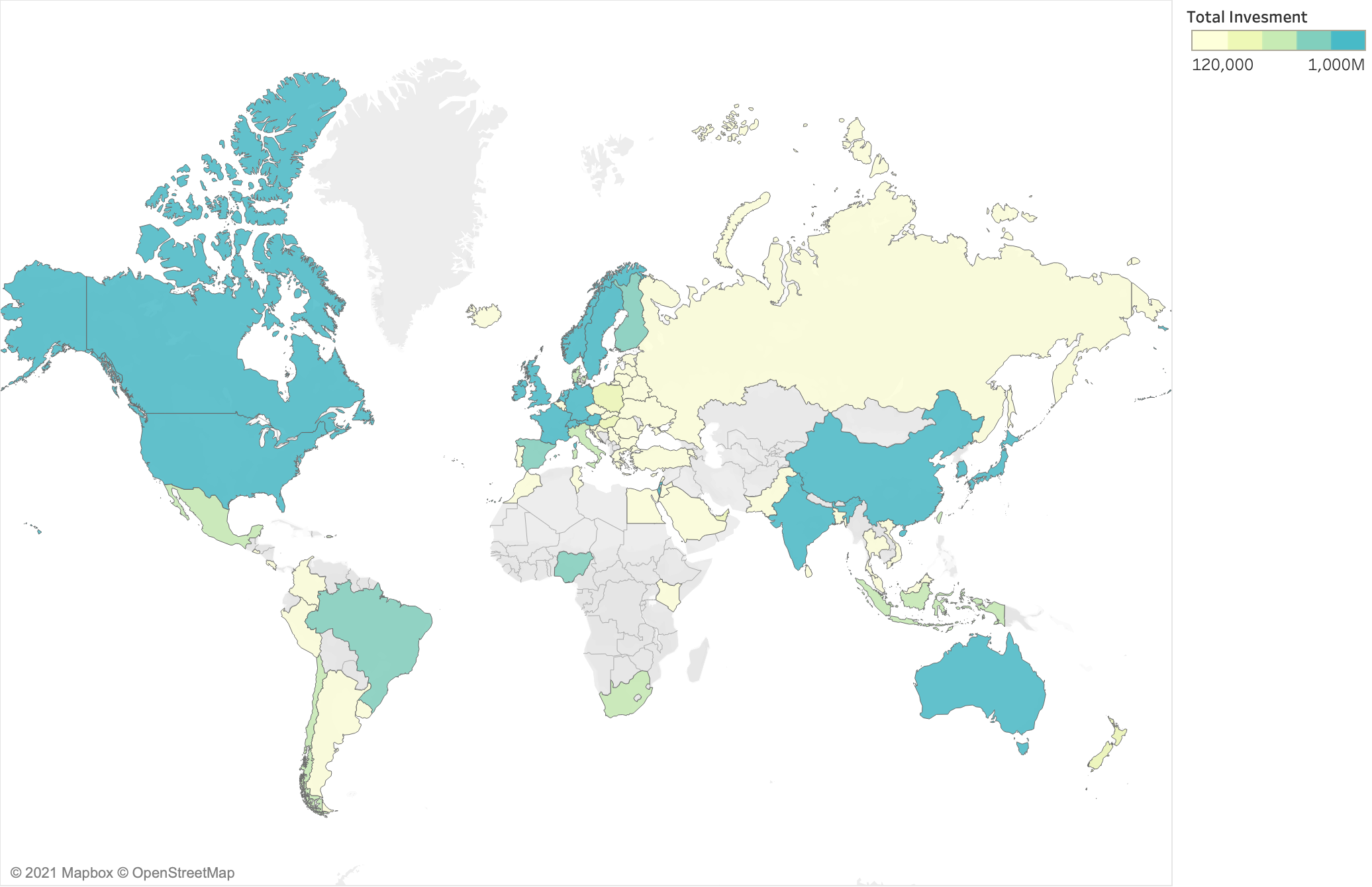}
    \caption{Global AI private investment, 2018-2020. The spectrum shows blue shade for countries with total AI investment over \$1B.}
    \label{fig:worldmap}
\end{figure}

\clearpage

\begin{figure}
    \centering
    \includegraphics[height=0.25\paperheight]{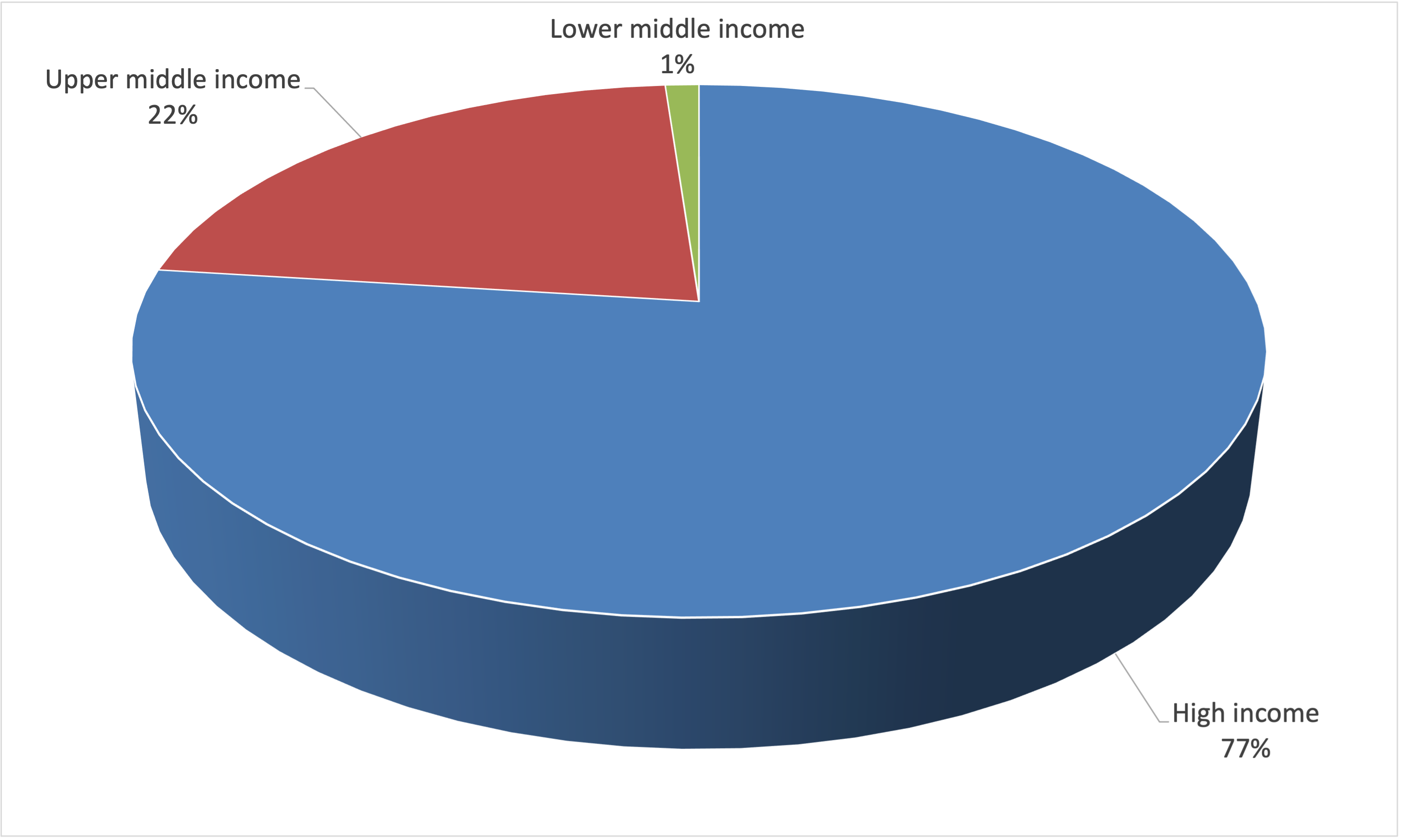}
    \caption{Percent of global AI private investment by World Bank income groups, 2019-2020.}
    \label{fig:income}
\end{figure}

\begin{figure}
    \centering
    \includegraphics[height=0.25\paperheight]{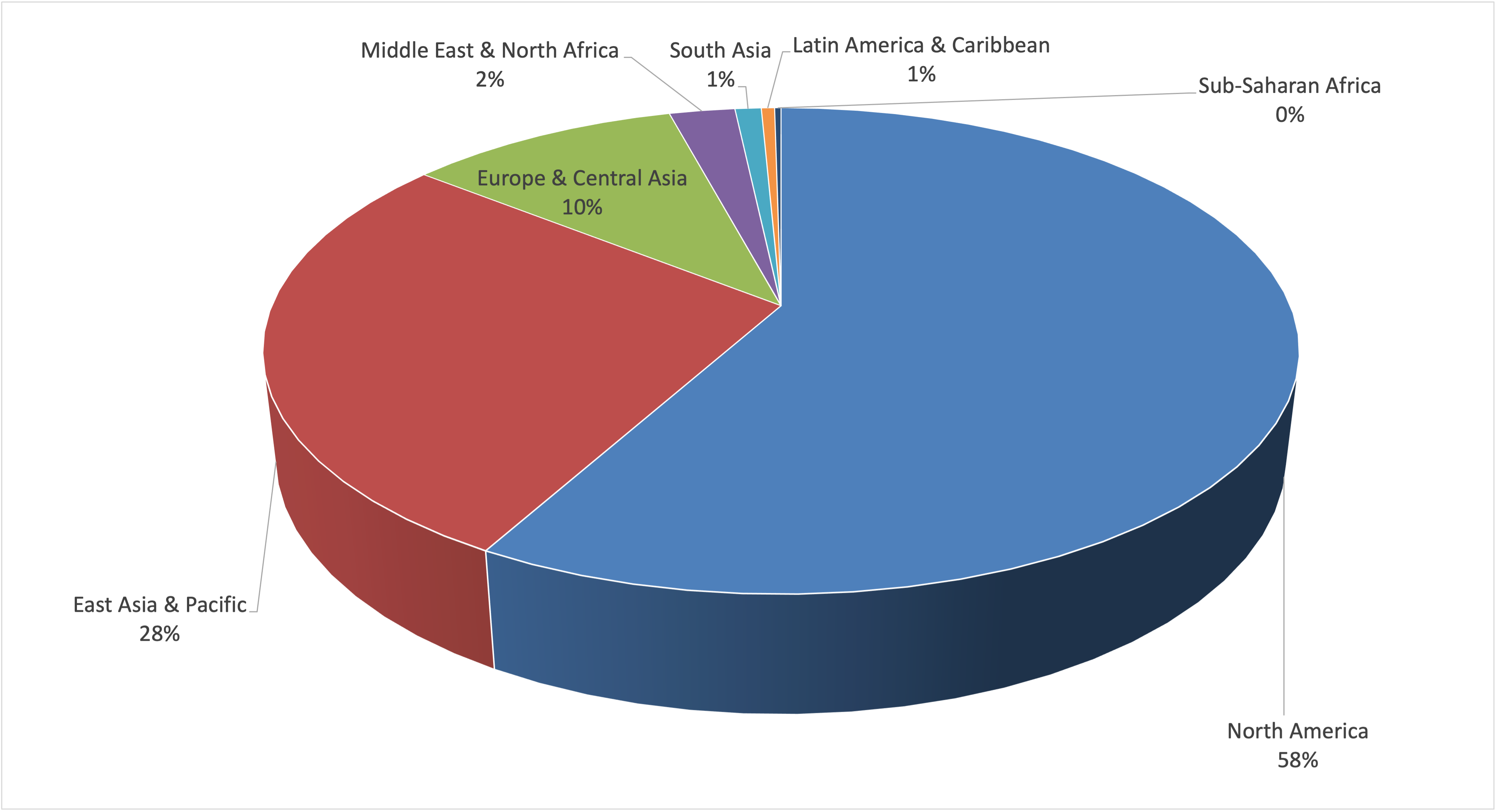}
    \caption{Percent of global AI private investment by World Bank regions, 2019-2020.}
    \label{fig:region}
\end{figure}

\clearpage

\begin{figure}
[ht]
    \centering
    \includegraphics[height=0.35\paperheight]{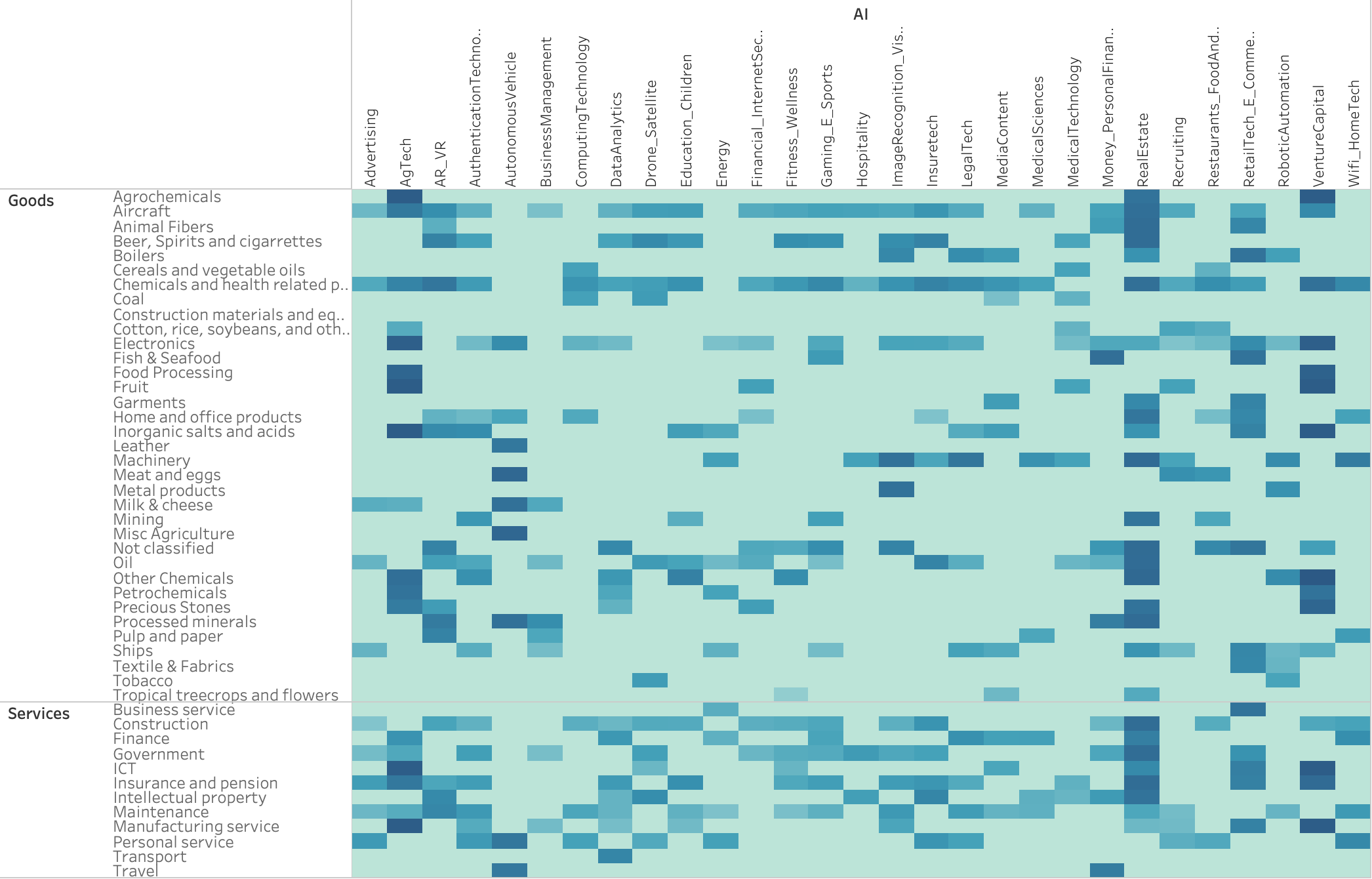}\newline
    \caption{Heatmap presentation of linkages from AI investment to goods and services export specialization. The darker shade of blue indicates statistically validated stronger linkages.}
    \label{fig:heatmap}
\end{figure}

\end{document}